\documentclass[longauth]{aa} 

\usepackage{graphicx}
\usepackage{txfonts}
\defcitealias{alo15}{Paper~I}

\begin{document}

\title{Variable stars in the VVV globular clusters.}

\subtitle{II. NGC~6441, NGC~6569, NGC~6626~(M~28), NGC~6656~(M~22), 2MASS-GC~02, and Terzan~10}

\author{Javier Alonso-Garc\'{i}a \inst{1,2}
  \and
  Leigh C. Smith\inst{3}
  \and
  M\'{a}rcio Catelan\inst{4,2}
  \and
  Dante Minniti\inst{5,6}
  \and
  Camila Navarrete\inst{7,2}
  \and
  Jura Borissova\inst{8,2}
  \and
  Julio A. Carballo-Bello\inst{9}
  \and
  Rodrigo Contreras Ramos\inst{4,2}
  \and
  Jos\'e G. Fern\'andez-Trincado\inst{10,11}
  \and
  Carlos E. Ferreira Lopes\inst{12}
  \and
  Felipe Gran\inst{4,2,7}
  \and
  Elisa R. Garro\inst{5}
  \and
  Doug Geisler\inst{13,14,15}
  \and
  Zhen Guo\inst{16}
  \and
  Maren Hempel\inst{5,17}
  \and
  Eamonn Kerins\inst{18}
  \and
  Philip W. Lucas\inst{16}
  \and
  Tali Palma\inst{19}
  \and
  Karla Pe\~na Ram\'irez\inst{1}
  \and
  Sebasti\'an Ram\'irez Alegr\'ia\inst{1}
  \and
  Roberto K. Saito\inst{20}
}

\institute{Centro de Astronom\'{i}a (CITEVA), Universidad de Antofagasta,
  Av. Angamos 601, Antofagasta, Chile\\
  \email{javier.alonso@uantof.cl}
  \and
  Millennium Institute of Astrophysics, Nuncio Monse\~nor Sotero Sanz 100,
  Of. 104, Providencia, Santiago, Chile
  \and
  Institute of Astronomy, University of Cambridge,
  Madingley Road, Cambridge, CB3 0HA, UK
  \and
  Instituto de Astrof\'{i}sica, Facultad de F\'isica,
  Pontificia Universidad Cat\'{o}lica de Chile,
  Av. Vicu\~na Mackenna 4860, 7820436 Macul, Santiago, Chile
  \and
  Departamento de Ciencias F\'isicas, Universidad Andr\'es Bello,
  Rep\'ublica 220, Santiago, Chile
  \and
  Vatican Observatory, Vatican City State V-00120, Italy
  \and
  European Southern Observatory, Alonso de C\'ordova 3107, Vitacura, Santiago, Chile
  \and
  Instituto de F\'isica y Astronom\'ia, Universidad de Valpara\'iso, Av. Gran Breta\~na 1111, Playa Ancha, Casilla 5030, Chile 
  \and
  Instituto de Alta Investigaci\'on, Sede Esmeralda, Universidad de Tarapac\'a, Av. Luis Emilio Recabarren 2477, Iquique, Chile
  \and
  Instituto de Astronom\'ia y Ciencias Planetarias, Universidad de Atacama, Copayapu 485, Copiap\'o, Chile
  \and
  Centro de Investigaci\'on en Astronom\'ia, Universidad Bernardo O Higgins, Avenida Viel 1497, Santiago, Chile
  \and
  Instituto Nacional de Pesquisas Espaciais (INPE), Av. dos Astronautas, 1758 - S\~ao Jos\'e dos Campos, SP 12227-010, Brazil
  \and
  Departamento de Astronom\'ia, Casilla 160-C, Universidad de Concepci\'on, Concepci\'on, Chile
  \and
  Instituto de Investigaci\'on Multidisciplinario en Ciencia y Tecnolog\'ia, Universidad de La Serena, Avenida Ra\'ul Bitrán S/N, La Serena, Chile
  \and
  Departamento de Astronom\'ia, Facultad de Ciencias, Universidad de La Serena, Av. Juan Cisternas 1200, La Serena, Chile
  \and
  Centre for Astrophysics, University of Hertfordshire, Hatfield AL10 9AB, UK
  \and
  Max-Planck-Institut f\"ur Astronomie, K\"onigstuhl 17, 69117 Heidelberg, Germany
  \and
  Jodrell Bank Centre for Astrophysics, Dept. of Physics and Astronomy, University of Manchester, Oxford Road, Manchester M13 9PL, UK
  \and
  Observatorio Astron\'{o}mico de C\'{o}rdoba, Universidad Nacional de C\'{o}rdoba, Laprida 854, X5000BGR, C\'{o}rdoba, Argentina
  \and
  Departamento de F\'{i}sica, Universidade Federal de Santa Catarina, Trindade 88040-900, Florian\'{o}polis, SC, Brazil
}

\date{Received ; accepted }

 
\abstract
   {The Galactic globular clusters (GGCs) located in the inner regions
     of the Milky Way suffer from high extinction that makes their
     observation challenging. High densities of field stars in their
     surroundings complicate their study even more. The {\em VISTA
     Variables in the Via Lactea} (VVV) survey provides a way to
     explore these GGCs in the near-infrared where extinction effects
     are highly diminished. }
   {We conduct a search for variable stars in several inner GGCs,
     taking advantage of the unique multi-epoch, wide-field,
     near-infrared photometry provided by the VVV survey. We are
     especially interested in detecting classical pulsators that will
     help us constrain the physical parameters of these GGCs. In this
     paper, the second of a series, we focus on NGC~6656~(M~22),
     NGC~6626~(M~28), NGC~6569, and NGC~6441; these four massive GGCs have
     known variable sources, but quite different metallicities. We
     also revisit 2MASS-GC~02 and Terzan~10, the two GGCs studied in
     the first paper of this series.}
   {We present an improved method and a new parameter that efficiently
     identify variable candidates in the GGCs. We also use the proper
     motions of those detected variable candidates and their positions
     in the sky and in the color-magnitude diagrams to assign
     membership to the GGCs.}
   {We identify and parametrize in the near-infrared numerous variable
     sources in the studied GGCs, cataloging tens of previously
     undetected variable stars. We recover many known classical pulsators in
     these clusters, including the vast majority of their fundamental
     mode RR~Lyrae. We use these pulsators to obtain distances and
     extinctions toward these objects. Recalibrated
     period-luminosity-metallicity relations for the RR~Lyrae bring
     the distances to these GGCs to a closer agreement with those
     reported by {\em Gaia}, except for NGC~6441, which is an uncommon
     Oosterhoff III GGC. Recovered proper motions for these GGCs also
     agree with those reported by {\em Gaia}, except for 2MASS-GC~02,
     the most reddened GGC in our sample, where the VVV near-infrared
     measurements provide a more accurate determination of its proper
     motions.}
   {}


   \maketitle
%

\section{Introduction}
\label{sec_intro}
Many Galactic globular clusters (GGCs) located in the inner regions of
the Milky Way (within 3 kpc from the Galactic center) still lack a
proper determination of their physical parameters. The analysis of the
color-magnitude diagrams (CMDs), which is the most common tool to
extract this information, is severely hampered when applied to these
objects. We need to add two specific issues more common in the inner
parts of the Galaxy to the usual problems we face when studying the
globular clusters of the outer Galaxy (e.g., high crowding, 
saturation by bright stars). The first issue is the presence of high
extinction and reddening, which usually change differentially over the
field of view of these GGCs. The second issue is the high density of
field stars that appear in the CMDs superimposed with the stellar
population of these GGCs, making it difficult to disentangle the field
from the cluster especially in the most poorly populated GGCs.

To diminish the effects of extinction, observations of these GGCs can
be carried out in the near-infrared. Extinction at these wavelengths
is significantly smaller than in the optical ($\mathcal{A}_K \sim 0.1
\mathcal{A}_V$; see Table~2 in \citealt{cat11}). To take full
advantage of this fact, the {\em VISTA Variables in the Via Lactea} (VVV)
survey \citep{min10,sai12b} observed the inner regions of the Galaxy
in the near-infrared in recent years.  VVV is a European Southern
Observatory (ESO) public survey that was conducted between 2010 and
2016, covering 560 sq. degrees of the Galactic bulge and an adjacent
region of the inner disk. Observations in five near-infrared filters
$ZYJHK_s$ were performed, and observations in $K_s$ of the whole
region were taken in multiple epochs, aiming to explore the presence
of variable stars and other variable phenomena in this area of the
sky.

There are 36 GGCs in the area covered by the VVV survey, according to
the 2010 version of the \citet{har96} catalog (from now on the Harris
catalog), along with tens of new candidates (e.g.,
\citealt{min17b,cam19,min19,gra19,pal19}; for a recent update, see
\citealt{bic19}). In a series of papers, we are exploring the variable
stars present in these GGCs, aiming to better characterize the cluster
in which they reside. Among them, RR~Lyrae stars are fundamental for
our purposes. Not only are these stars quite common in (many) globular
clusters, but their period-luminosity (PL) relation, especially tight
in the near-infrared \citep{lon86,cat04}, makes them excellent
standard candles that allow us to accurately infer their distances and
extinctions; we showed this for 2MASS-GC~02 and Terzan~10 in
\citet{alo15}, which we refer to from now on as
\citetalias{alo15}. The GGCs are clumped into two main groups
\citep{oos39,cat09a,smi11}, according to the characteristics of the
fundamental-mode RR~Lyrae (RRab) that they contain: the Oosterhoff~I
group shows RRab stars with shorter periods ($\langle P_{\rm ab}
\rangle \sim 0.55$~days), while the Oosterhoff II have RRab stars with
longer periods ($\langle P_{\rm ab} \rangle \sim 0.64$~days). Between
these two groups, there is an almost empty region called the
Oosterhoff gap, at $\langle P_{\rm ab} \rangle \sim
0.60\pm0.02$~days. Oosterhoff~II GGCs also tend to be more metal-poor
than Oosterhoff~I GGCs and to have a higher ratio of first-overtone
RR~Lyrae (RRc) to RRab stars. A couple of GGCs containing $\langle
P_{\rm ab} \rangle$ too long for their high metallicities have been
classified as Oosterhoff~III GGCs \citep{pri00}.

\begin{table*}
  \caption{Positions and physical parameters of the target clusters.}
  \label{tab_gcinput}
  \centering
  \begin{tabular}{ccccccccc}
    \hline\hline
    Cluster & $\alpha$(J2000) & $\delta$(J2000) & $l$ & $b$ & $[{\rm Fe/H}]$ & $M_V$ & c & $r_{t}$ \\
    & (h:m:s) & (d:m:s) & (deg) & (deg) & (dex)& (mag) & & (arcmin) \\
    \hline
    NGC~6441 & 17:50:13 & -37:03:04 & 353.53 & -5.01 & -0.46 & -9.63 & 1.74 & 7.14 \\
    Terzan~10 & 18:02:58 & -26:04:00 & 4.42 & -1.86 & -1.59 & -6.35 & 0.75 & 5.06 \\
    2MASS-GC~02 & 18:09:37 & -20:46:44 & 9.78 & -0.62 & -1.08 & -4.86 & 0.95 & 4.90 \\
    NGC~6569 & 18:13:39 & -31:49:35 & 0.48 & -6.68 & -0.76 & -8.28 & 1.31 & 7.15 \\
    NGC~6626 (M~28) & 18:24:33 & -24:52:12 & 7.80 & -5.58 & -1.32 & -8.16 & 1.67 & 11.22 \\
    NGC~6656 (M~22) & 18:36:24 & -23:54:12 & 9.89 & -7.55 & -1.70 & -8.50 & 1.38 & 31.90 \\
    \hline
  \end{tabular}
  \tablefoot{Equatorial coordinates are taken from \citet{bic19}. Iron
    content $[{\rm Fe/H}]$, absolute integrated visual magnitude
    $M_V$, concentration $c= \log(r_t/r_c)$, and tidal radii $r_{t}$
    are according to the 2010 version of the \citet{har96} catalog,
    except for Terzan~10, whose $[{\rm Fe/H}]$ value is provided by
    Geisler et al. (in prep.), as detailed in Sect.~\ref{sec_gc2var}.}
\end{table*}

In this second paper of the series, we focus on several well-known
GGCs located in the VVV footprint: NGC~6441, NGC~6569,
NGC~6626~(M~28), and NGC~6656~(M~22). These GGCs show a significant
range in their metallicities (see Table~\ref{tab_gcinput}) and in
their Oosterhoff classification. They differ from those studied in
\citetalias{alo15} because they lie in regions in which extinction is
lower, although still high for outer GGCs standards. These GGCs are
also better populated than the GGCs studied in \citetalias{alo15}, and
they lie in fields in which the stellar background densities are
lower. Finally, they possess recent distance estimations inferred from
{\em Gaia} data \citep{bau19}, and they contain significant numbers of
variable stars in their fields present in the most recent version of
the \citet{cle01} catalog of variable stars in the GGCs (from now on,
the Clement catalog) and in the collection of variable stars in the
inner Milky Way by \citet{sos16,sos17,sos19} from the Optical
Gravitational Lensing Experiment (OGLE). Therefore, by drawing a
comparison with this previous literature, we aim to examine the
reliability of our methods to detect variable stars and to infer the
physical parameters (distance, extinction, and proper motion [PM])
for their GGCs. While achieving this, we also provide a look at the
variable stars of these GGCs from their innermost regions to their
outskirts at near-infrared wavelengths for the first
time. Additionally, we also revisit 2MASS-GC~02 and Terzan~10, the
GGCs from \citetalias{alo15}, to check the advantages and drawbacks of
our updated approach to detect variable sources.

The paper is divided into the following sections: In
Sect.~\ref{sec_obs}, we describe our database; in Sect.~\ref{sec_var}
we develop our improved method to detect variable stars; in
Sect.~\ref{sec_m22}, \ref{sec_gc1}, and \ref{sec_gc2} we implement
this method firstly on NGC~6656~(M~22), then on NGC~6441, NGC~6569,
NGC~6626~(M~28), and finally on 2MASS-GC~02 and Terzan~10, providing a
description of the variable star candidates we found in their fields;
we also select potential members of each cluster according to their PM
and their positions in the CMD, and use these sources to obtain the
distances and extinctions toward these GGCs; finally, in
Sect.~\ref{sec_con} we summarize our conclusions.

\section{Observations and datasets}
\label{sec_obs}
In our analysis, we used observations from VVV \citep{min10}, one of
the six original ESO public surveys conducted with the 4.1m VISTA
telescope in the Cerro Paranal Observatory. The camera installed in
the telescope provides wide-field images of 1.6 square degrees of the
sky, with gaps due to the separation of the 16 chips in the
camera. The VVV survey uses five near-infrared filters ($Z$, $Y$, $J$,
$H$, and $K_s$) and its footprint for the Galactic bulge region is
divided into 196 individual fields. The observing strategy for every
VVV field, detailed in \citet{sai12b}, consists in taking two
consecutive slightly dithered images of the sky in a given filter
which, when later on combined into a so-called stacked image, allow
the correction of cosmetic defects from the different chips. Along
with this pattern, a mosaic of six consecutive pointings is taken for
every field and filter to provide a contiguous coverage of the
observed field, covering the gaps among the chips in the camera. Every
field is observed at least twice in $Z$, $Y$, $J$, and $H$, and at
least 70 times in the $K_s$ filter. The $K_s$ observations in every
epoch have a median exposure time of 16 seconds \citep{sai12b} and
were taken in a nonuniform, space-varying cadence \citep{dek19}.

Our analysis is based on the VVV photometry and astrometry provided by
VIRAC2, an updated version of the VVV Infrared Astrometric Catalogue
\citep[VIRAC;][]{smi18}. A complete description of the new features of
VIRAC2 will be given in an upcoming dedicated paper (Smith et al., in
prep.). Suffice it to say here that this version of VIRAC uses DoPHOT
\citep{sch93,alo12} to extract the point-spread function photometry
from the VVV stacked images, significantly reducing the missing
sources and increasing the completeness of the sample, especially in
the highly crowded GGCs \citep{alo18}.  Photometric zero points for
each observation were measured by globally minimizing the
error-normalized offsets between multiple detections of individual
sources and offsets from 2MASS (transformed to the VISTA photometric
system as per \citealt{gonfer18}) where available. A further
calibration was applied to remove spatially coherent residual
structure and match the photometric uncertainties to residual
scatter. This way the calibration offsets reported in \citet{haj20}
are effectively resolved. The VIRAC2 catalog provides us with the VVV
mean magnitudes, PMs, and near-infrared light curves for all the
detected sources.

\section{Variability analysis}
\label{sec_var}
We modified the variability analysis presented in \citetalias{alo15}
to optimize for its use with VIRAC2 on the detection and proper
characterization of classical pulsators. These are the variable stars
that we are more interested in detecting because they can be used as
standard candles. According to our experience, Cepheids and RRab stars
are the classical pulsators that stand out the most in the
near-infrared VVV observations. Even though their light curves are
more sinusoidal and their amplitudes are smaller in the near-infrared
than at optical wavelengths \citep{ang14,cat15}, the features of their
light curves still provide ways to properly identify them within the
VVV data, as shown in \citetalias{alo15}. The smaller amplitudes of
the RRc stars complicate their identification as variable
sources. Even when they are identified as variable stars, their almost
sinusoidal light curves make it difficult to conclusively assign them
to this class using only their VVV near-infrared
photometry. Therefore, the analysis described in this section aims to
mainly identify most of the Cepheids and RRab stars with good-quality
light curves in the VVV GGCs fields. Some RRc stars and eclipsing
binaries are also identified, as shown in the next sections.

\begin{figure}
  \centering
  \includegraphics[scale=0.65]{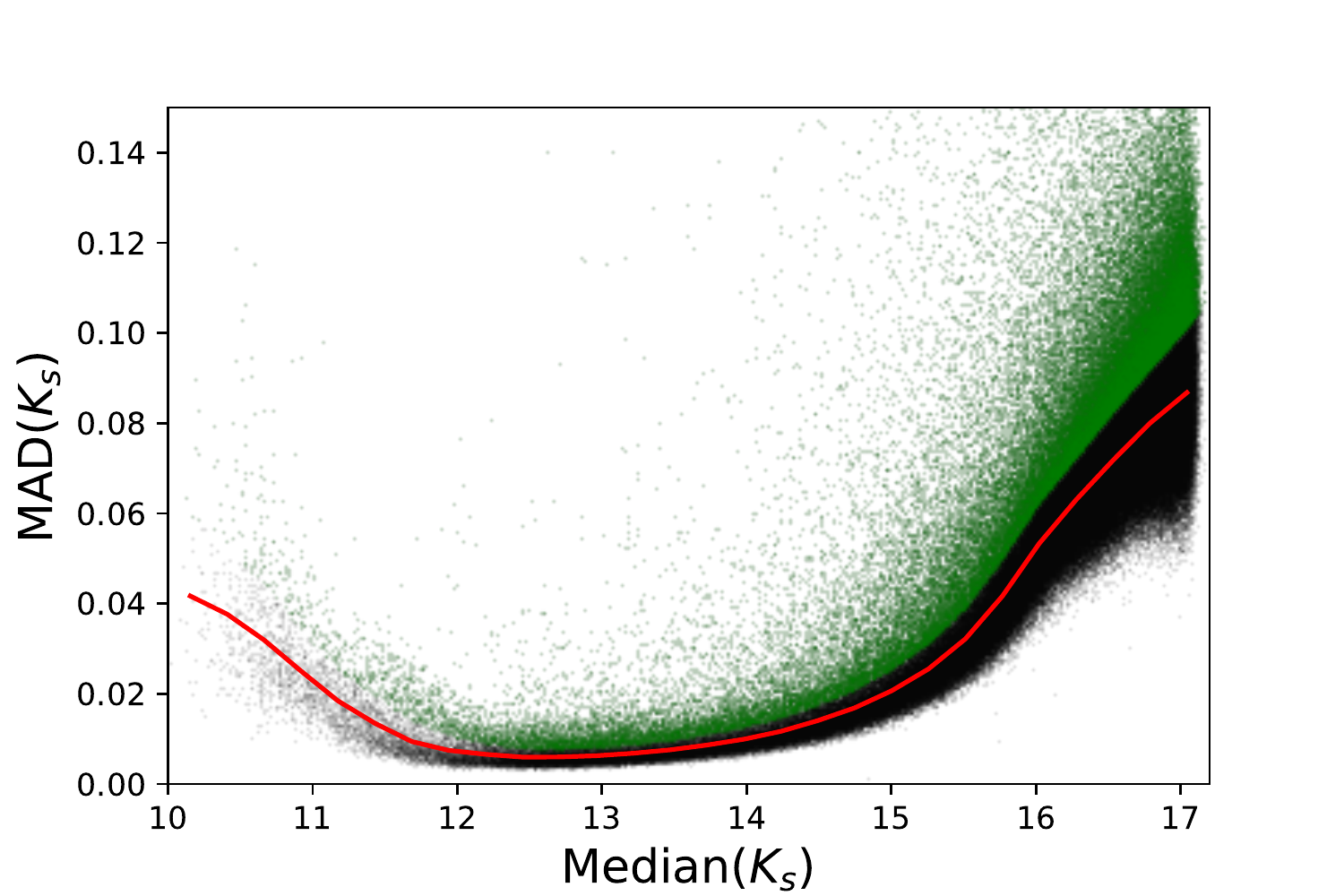}
  \caption{MAD vs. median of the $K_s$ magnitudes for the different
    epochs with VVV detections for the stars in M~22, serving as an
    example of the preliminary selection to detect variable
    candidates. The green dots show the stars we keep for the next
    step of the analysis. They are located $1\sigma$ above the median
    of the distribution, shown by the red line.}
  \label{fig_mad}
\end{figure}

The first step in our analysis was to take all sources from VIRAC2
that lie inside the tidal radius of the selected GGCs (see
Table~\ref{tab_gcinput}). After that, we selected stars that were
measured in at least half of the $K_s$ observations that were taken
for their region of sky. That way we disregarded spurious or very low
signal-to-noise detections. The next step in our analysis looked at
the distribution of the median absolute deviation (MAD) of the $K_s$
magnitudes in the different epochs for the stars in a given cluster,
as a function of its median $K_s$ magnitude. As classical pulsators
show variations along all its phased light curve, this produces a
higher MAD than for a non-variable source at a given magnitude. For
other types of variable stars such as eclipsing binaries with short
and/or not well-sampled eclipses, the MAD may not be such a good
indicator of variability. We calculated the median and the standard
deviation of the MAD as a function of the median $K_s$ magnitude. For
the next step of our analysis we kept only sources that are more than
$1\sigma$ away from the median of the MAD (see Fig.~\ref{fig_mad}). We
noticed from some preliminary tests with previously known variables in
our sample of GGCs that for stars under this cutoff we were unable to
detect classical pulsators with good-quality light curves. By applying
this cutoff, we reduced the number of stars to be analyzed to $\sim
10\%$ of the original sample.

\begin{figure}
  \centering
  \includegraphics[scale=0.5]{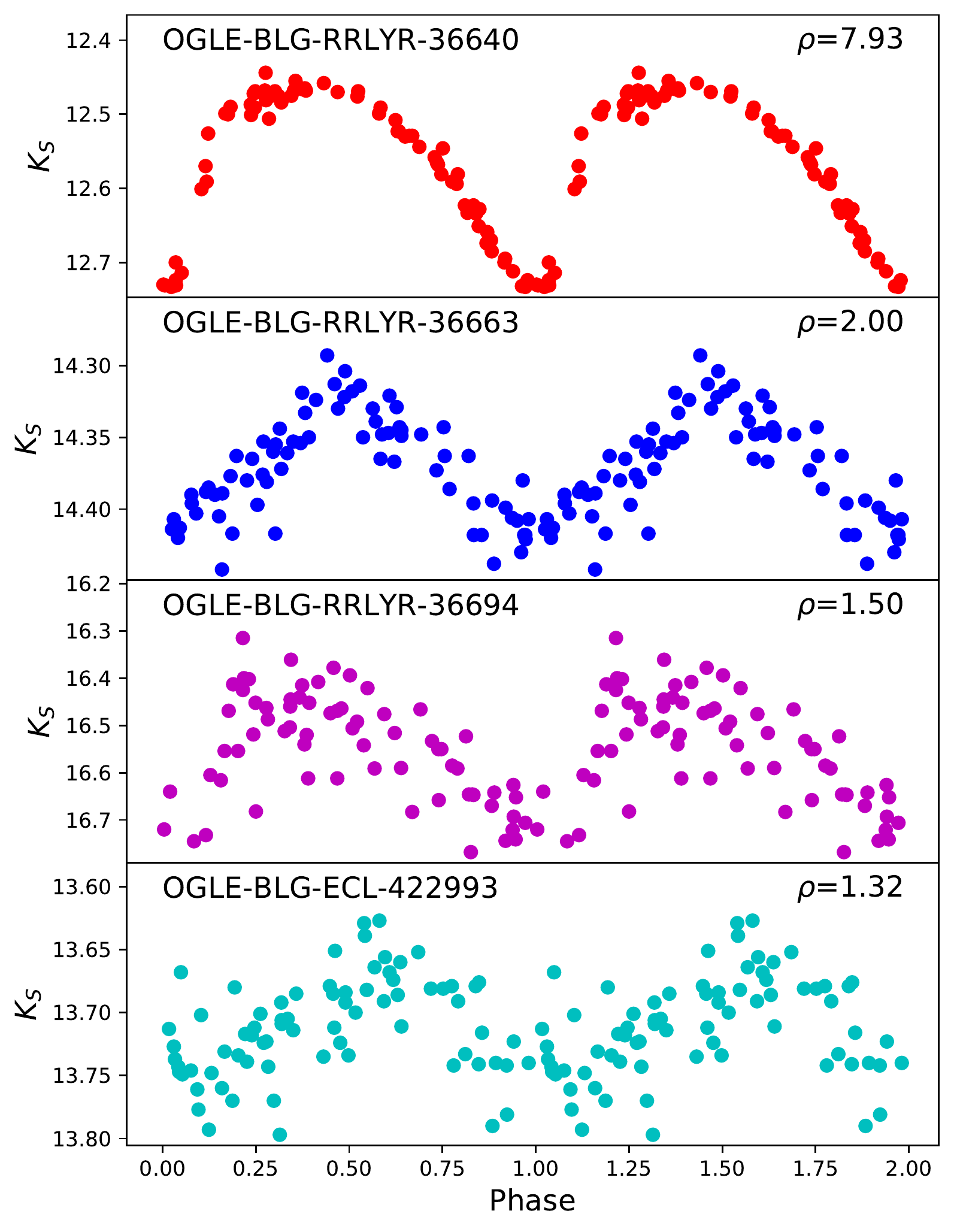}
  \caption{VVV $K_s$ phase-folded light curves for some variable stars
    in M~22 in common with the OGLE catalog, as examples of how their
    quality changes with the $\rho$ parameter defined in the
    text. From top to bottom, light curves are ordered according to
    their decreasing $\rho$ value. The upper two panels, with the
    higher $\rho$ values, represent clear identifications of the
    variable nature of these stars according to their VVV light
    curves. The third panel, where $\rho=1.50$, represents a detection
    just in our cutoff point. In the lower panel, the poor quality of
    the VVV light curve makes it difficult to separate this true
    variable source from other false triggers with similar $\rho$
    values, rendering their identification as a real variable
    problematic.}
  \label{fig_varselect}
\end{figure}

We then looked for periodicity in the signal of every source we
kept. In order to do that, first we submitted the light curve of the
remaining sources to a generalized Lomb-Scargle (GLS) analysis
\citep{zec09}. We analyzed the interval of frequencies $[0.01-10]$
days$^{-1}$, which covers all RR~Lyrae and Cepheids. As in
\citetalias{alo15}, we eliminated sources with significant gaps in a
given section of their folded light curve and aliased sources with
values for frequency close to integers of days$^{-1}$. Using the best
period the GLS algorithm provided, we fitted a Fourier series to the
folded light curve. We masked those epochs that departed more than
$3\sigma$ from the Fourier fit. We averaged values from the same
mosaic sequence (see Sect.~\ref{sec_obs}) for a more accurate sampling
of the light curve. Since the time taken to observe these sequences
for a given VVV field and filter is much shorter than the period of
the variable sources we are measuring, we can consider observations in
one of these sequences to have been taken at the same epoch. We then
recalculated the period of the variable candidate using the GLS
algorithm, and repeated the described sequence iteratively until we
reached a convergence in the period. After this, for every source we
calculated $\rho$, the ratio of the standard deviation of the
distribution of $K_s$ mosaic-averaged magnitudes from the different
epochs available for that source to the standard deviation from its
Fourier fit. The ratio $\rho$ between these two standard deviations
provides us with a good approximation to a by-eye identification and
quality assignment for variable sources (see
Fig.~\ref{fig_varselect}). After some visual inspection, we adopted a
value of $\rho=1.50$ as our cutoff point for the variable stars in the
GGCs reported in this study, except for Terzan~10 in which we used
$\rho=1.30$ as our cutoff point owing to the higher number of epochs
available for this cluster. A quick visual inspection of the shape of
the light curves above this cutoff value allowed us to reject a few
false positives, mainly owing to blending from two sources in the same
light curve.

For all the remaining variable candidates, we proceeded to obtain
their observational parameters. As in \citetalias{alo15}, the apparent
$K_s$-band equilibrium brightnesses of the stars $\langle K_s \rangle$
were estimated by the intensity-averaged magnitudes of the stars,
computed from the Fourier fits to the light curves, and the total
amplitudes of the light curves $A_{K_s}$ were computed from the
Fourier fits as well. We have observations for filters $Z$, $Y$, $J$,
and $H$ only in approximately two epochs per filter. To calculate the
$\lambda-K_s$ apparent colors, we obtained the $K_s$ magnitude from
the Fourier fit of the $K_s$ light curve at the same phase where the
measurements for the other filters were taken, and after that, we took
the mean of the colors from the few different epochs available for a
given filter. Finally, we proceeded to examine the light curves to
assign a variable type to the different candidates. As mentioned in
\citetalias{alo15}, this eyeball classification is complicated by the
fact that the near-infrared $K_s$ light curves usually contain fewer
outstanding features than in the optical. Although we could reliably
classify Cepheids, RRab stars, and some eclipsing binaries, there were
a significant number of candidates that we left with their variable
type as undecided. For stars in common with the other catalogs and
variable types difficult to characterize in the near-infrared (e.g.,
RRc stars, W~UMa-type [EW] eclipsing binaries), we kept the
classification from their optical light curves. For stars classified
as eclipsing binaries, we doubled the periods reported by our
algorithms because they do not separate primary and secondary eclipses
in these variable stars.
  
\section{NGC~6656 (M~22)}
\label{sec_m22}
NGC~6656 (M~22) is the most metal-poor cluster from the sample of VVV
GGCs we are considering in this paper. Its high brightness and
moderate concentration when compared with other VVV GGCs (see
Table~\ref{tab_gcinput} for its physical characteristics), along with
its relative proximity, makes NGC~6656 the GGC that subtends the
largest sky area in the original VVV footprint, rivaled only by
FSR~1758 \citep{bar19}, observed by the VVV eXtended survey
\citep[VVVX;][]{min18d}.

\subsection{Variables in the cluster area}
\label{sec_m22var}
There have been a significant number of studies looking for variables
in this Oosterhoff~II GGC, dating back to those by \citet{bai02},
\citet{sha27}, and \citet{saw44}, up to the most recent works by
\citet{kun13} and \citet{roz17}. Ours is the first study in the
near-infrared that covers the whole cluster, from its very center out
to its tidal radius. Following the steps described in
Sect.~\ref{sec_var}, we identified 439 variable candidates in the
field of this GGC. We present their physical properties in
Table~\ref{tab_m22var}. There are 142 variable sources reported in the
Clement catalog for M~22, while we found 604 variable stars in the
OGLE catalog inside the tidal radius of M~22 and there are 360
variable sources in the catalog presented by \citet{roz17}. As shown
in the comparison in Table~\ref{tab_m22var}, most variable stars in
our catalog are also present in these other catalogs, but there are
155 variable candidates that were not previously reported.

\begin{table*}
  \caption{Properties of the variable candidates in NGC~6656 (M~22).}
  \label{tab_m22var}
  \centering
  \resizebox{\textwidth}{!}{
  \begin{tabular}{cccccccccccccccccc}
    \hline\hline
    ID & $ID_{Clement}$ & $ID_{OGLE}$ & $ID_{Ros17}$ & $\alpha$ (J2000) & ${\delta}$ (J2000) & Distance\tablefootmark{a} & Period & $A_{K_s}$ & $\langle K_s \rangle$ & $Z-K_s$ & $Y-K_s$ & $J-K_s$ & $H-K_s$ & $\mu_{\alpha^\ast}$ & $\mu_{\delta}$ & Member\tablefootmark{b} & Type \\
    & & (OGLE-BLG-) & & (h:m:s) & (d:m:s) & (arcmin) & (days) & (mag) & (mag)& (mag)& (mag)& (mag) & (mag) & (mas/yr) & (mas/yr) & & \\ 
    \hline
C1 & KT-55 & -- & KT-55 & 18:36:23.25 & -23:53:58.1 & 0.29 & 0.658743 & 0.255 & 12.047 & 0.699 & 0.538 & 0.386 & 0.085 & 9.322 & -5.336 & Yes & RRab\\
C2 & V24 & T2CEP-0927 & 24 & 18:36:21.80 & -23:54:13.4 & 0.5 & 1.714805 & 0.284 & 11.104 & 0.988 & 0.774 & -- & 0.75 & 10.54 & -4.359 & Yes & Cep\\
C3 & V23 & RRLYR-36672 & 23 & 18:36:23.04 & -23:54:41.7 & 0.54 & 0.551593 & 0.307 & 12.253 & 0.911 & 0.704 & 0.412 & 0.104 & 10.056 & -5.662 & Yes & RRab\\
C4 & V1 & RRLYR-36670 & 1 & 18:36:19.57 & -23:54:33.0 & 1.07 & 0.615536 & 0.309 & 12.12 & 0.883 & 0.669 & 0.495 & 0.115 & 8.909 & -5.574 & Yes & RRab\\
C5 & -- & -- & -- & 18:36:25.20 & -23:53:04.8 & 1.15 & 0.105017 & 0.311 & 15.689 & -- & 0.382 & 0.236 & 0.028 & 16.141 & 1.402 & Yes & ?\\
C6 & V21 & RRLYR-36675 & 21 & 18:36:25.76 & -23:52:58.2 & 1.29 & 0.327135 & 0.08 & 12.441 & 0.622 & 0.448 & 0.368 & 0.096 & 9.948 & -5.237 & Yes & RRc\\
C7 & V4 & RRLYR-36673 & 4 & 18:36:23.37 & -23:55:29.4 & 1.3 & 0.716398 & 0.28 & 11.959 & 0.913 & 0.695 & 0.498 & 0.115 & 10.284 & -6.561 & Yes & RRab\\
C8 & V12 & RRLYR-36674 & 12 & 18:36:23.77 & -23:55:39.1 & 1.45 & 0.322624 & 0.099 & 12.451 & 0.682 & 0.492 & 0.335 & 0.069 & 10.372 & -5.292 & Yes & RRc\\
C9 & KT-14 & -- & KT-14 & 18:36:30.69 & -23:53:54.1 & 1.56 & 0.373642 & 0.081 & 12.291 & 0.735 & 0.534 & 0.355 & 0.091 & 10.267 & -7.793 & Yes & RRc\\
C10 & KT-12 & RRLYR-36679 & KT-12 & 18:36:30.94 & -23:53:49.1 & 1.63 & 0.443619 & 0.222 & 14.572 & 0.731 & 0.567 & 0.496 & 0.17 & -4.454 & -3.644 & No & RRab\\
    \hline
  \end{tabular}}
\tablefoot{This table is available in its entirety in electronic form
  at the Centre de Donn\'ees astronomiques de Strasbourg (CDS). A
  portion is shown here for guidance regarding its form and
  content.\\ \tablefoottext{a}{Projected distance to the cluster
    center} \tablefoottext{b}{Cluster membership according to criteria
    explained in Sect.~\ref{sec_m22dis}} }
\end{table*}

It is interesting to highlight that we recovered the vast majority of
the previously reported RRab present in the cluster region. From the
39 RRab stars reported in the literature, there are only 2
(OGLE-BLG-RRLYR-36695 and OGLE-BLG-RRLYR-64197) and 1 possible RRab
(U37) that we were not able to detect as variable sources following
the steps described in Sect.~\ref{sec_var}. According to the Clement
catalog, and judging from their dim optical magnitudes, these
undetected RRab stars are not cluster members, but background objects
whose low signal-to-noise ratios in our VVV data may hamper our
ability to classify them as variables. So our method is able to
recover all the previously known RRab sources that belong to M~22. On
the other hand, we report the discovery of 5 possible uncharted RRab
stars in the cluster region (C112, C150, C196, C385, and C424),
although they are highly unlikely to be cluster members judging by
their dim near-infrared magnitudes and their PMs (see
Sect.~\ref{sec_m22pm}).

As mentioned in Sect.~\ref{sec_var}, detecting a lower proportion of
RRc stars with good-quality light curve is expected as a consequence
of their smaller amplitudes. We note that we missed 10 of the 35
previously reported RRc sources. Among those, only 4 (Ku-2, KT-16,
KT-26, and KT-37) are cluster members, according to the Clement
catalog. However, we report 1 new RRc candidate, C273, although its
dim near-infrared magnitudes and PM (see Sect.~\ref{sec_m22pm}) make
it highly unlikely for it to be a cluster member as well. We note as
well that we only recovered C2 (V24), the dimmest of the two Cepheid
stars reported in the literature, which we attribute to the brightest
one (V11) being saturated in our near-infrared data. It is also
interesting to highlight that most of the 155 newly found variable
stars are left as unknown types. We were only able to classify 5 RRab,
1 RRc and 16 eclipsing binaries with very clear eclipses. Finally, we
note 3 stars that are classified differently in the various catalogs
that we checked: C6 (V21), which appears as RRab or RRc; C30 (Ku-4),
which appears as eclipsing binary or RRc; and C56, which appears as
eclipsing binary or semiregular. According to their near-infrared
light curves, we classified C6 and C30 as RRc stars, and C56 as an
eclipsing binary.

\subsection{Proper motion, color-magnitude diagram, and cluster membership}
\label{sec_m22pm}
\begin{figure*}
  \centering
  \begin{tabular}{cc} 
  \includegraphics[scale=0.5]{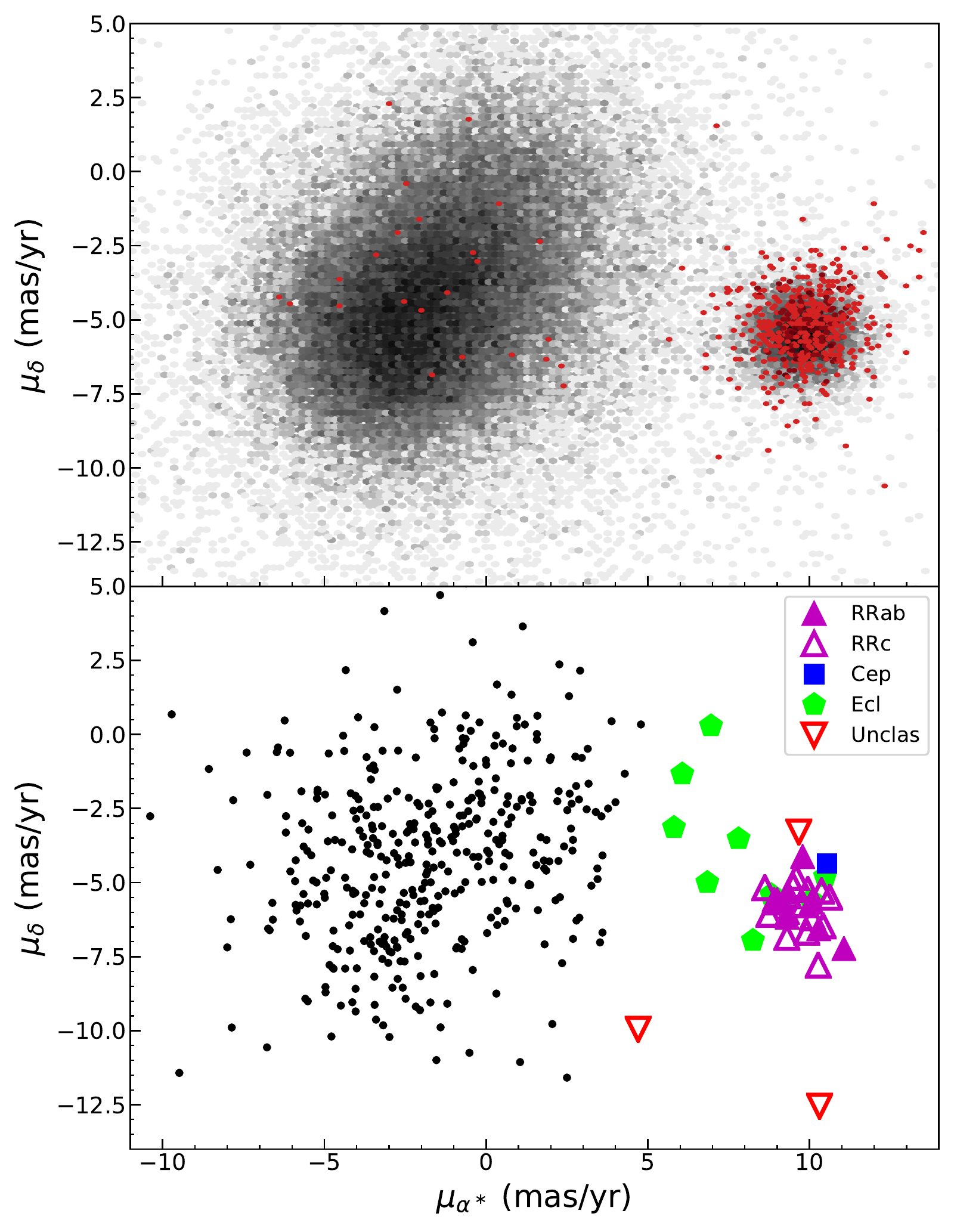}
  \includegraphics[scale=0.5]{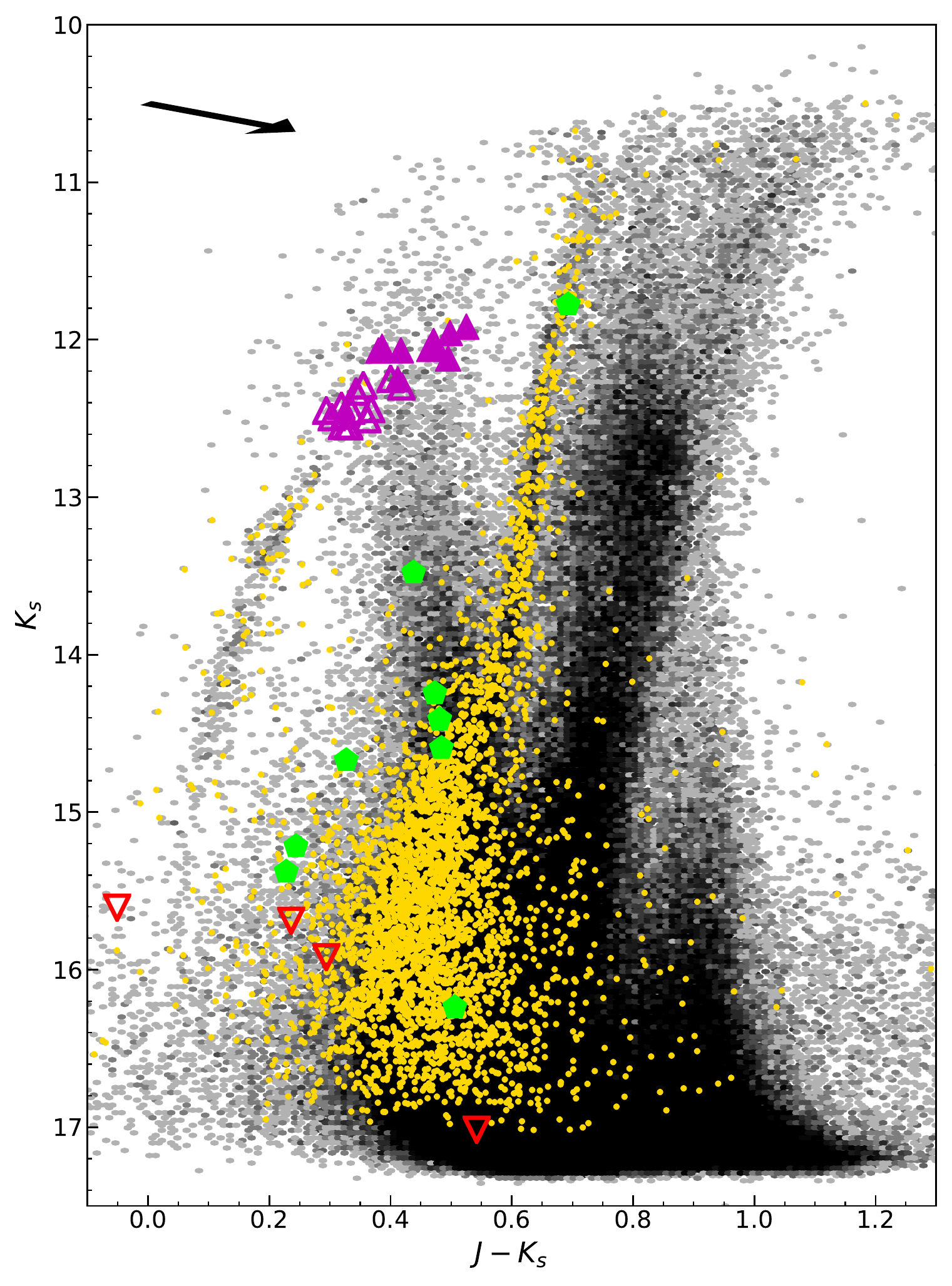}
  \end{tabular}
  \caption{Upper left panel: PMs of the stars in the M~22 region with
    $\sigma\leq1.0$ mas/yr. Higher transparencies represent lower
    densities of stars. Overplotted in red are the PMs of the stars
    located in the innermost regions ($r\leq1.0'$) of this GGC. Lower
    left panel: PMs of the detected variable stars in the field covered
    by M~22. Those stars selected as cluster members by our kNN
    classifier are shown according to their variability type: the
    solid magenta triangles indicate RRab stars, empty magenta
    triangles are RRc stars, blue squares are Cepheids, green
    pentagons are eclipsing binaries, and inverted red triangles are
    unclassified variables. Right panel: $J-K_s$ vs. $K_s$
    near-infrared CMD of the stars in the M~22 field. Higher
    transparencies represent lower densities of stars. Overplotted in
    yellow are the stars located in the innermost regions
    ($r\leq1.0'$) of this GGC that were selected as candidates by our
    kNN classifier. The cluster member variables from the lower left
    panel are also overplotted. The arrow in the upper left corner
    shows the selected reddening vector.}
  \label{fig_m22}
\end{figure*}

M~22 is one of the GGCs with the highest PM \citep{bau19}. Using the
multi-epoch VVV observations available at the time, \citet{lib15}
managed to separate the cluster stars from field stars using PMs
selection. One of the main features of the VIRAC2 database is to
provide accurate PMs for the stars in the VVV Galactic bulge
footprint. If we select stars in the field of M~22 with
well-determined VIRAC2 PMs ($\sigma_{\mu_{\alpha^\ast}}<1$ mas/yr,
$\sigma_{{\mu}_{\delta}}<1$ mas/yr), we can observe a clear separation
between two main distributions (see upper left panel of
Fig.~\ref{fig_m22}). The closer we move to the center of the cluster,
the higher the probability for the star to belong to the cluster
\citep{alo11}. Hence, selecting stars close to the cluster center
($r\leq1'$) allows for a clearer identification of the distribution of
stars that belong to M~22, as we can observe from the red dots in the
upper left panel of Fig.~\ref{fig_m22}. In order to define a criterion
for the stars to belong to the cluster based on the PM of our sample,
we used a k-nearest neighbors (kNN) classification \citep{has09}. To
train the kNN classifier, we selected as test stars belonging to the
cluster those with distances $r\leq1'$, and as test stars belonging to
the field an equal number of stars, but we selected the stars in our
sample with the farthest distance from the cluster center. The number
of nearest neighbors used by the kNN classifier is one tenth of the
total test sample. Selecting the innermost ($r\leq1'$) stars that
comply with our membership criteria allows us not only to properly
identify cluster stars in the CMD (see right panel of
Fig.~\ref{fig_m22}), but also to accurately define the PM of M~22 as
the mean of the PMs of those selected stars (see Table~\ref{tab_gcpm}),
which closely agrees with the PM obtained for this GGC from {\em Gaia}
Data Release 2 \citep[DR2;][]{bau19}.

\begin{table*}
  \caption{Proper motions of the target clusters.}
  \label{tab_gcpm}
  \centering
  \begin{tabular}{ccccc}
    \hline\hline
    Cluster & $\mu_{\alpha^\ast}$ & $\mu_{\delta}$ & $\mu_{\alpha^\ast Gaia}$ & $\mu_{\delta Gaia}$ \\
     & (mas/yr) & (mas/yr) & (mas/yr) & (mas/yr) \\
    \hline
    NGC~6441 & $-2.4\pm0.8$ & $-5.6\pm1.0$ &$-2.51\pm0.03$ & $-5.32\pm0.03$ \\
    Terzan~10 & $-6.8\pm1.2$ & $-2.5\pm1.3$ & $-6.91\pm0.06$ & $-2.40\pm0.05$ \\ 
    2MASS-GC~02 & $4.0\pm0.9$ & $-4.7\pm0.8$ & $-1.97\pm0.16$ & $-3.72\pm0.15$ \\
    NGC~6569 & $-4.1\pm0.8$ & $-7.3\pm0.8$ & $-4.13\pm0.02$ & $-7.26\pm0.02$ \\
    NGC~6626 (M~28) & $-0.3\pm2.0$ & $-9.0\pm1.5$ & $-0.33\pm0.02$ & $-8.92\pm0.02$ \\
    NGC~6656 (M~22) & $9.9\pm1.1$ & $-5.2\pm1.2$ & $9.82\pm0.01$ & $-5.54\pm0.01$ \\
    \hline
  \end{tabular}
\end{table*}

\begin{figure*}
  \centering
  \includegraphics[scale=0.85]{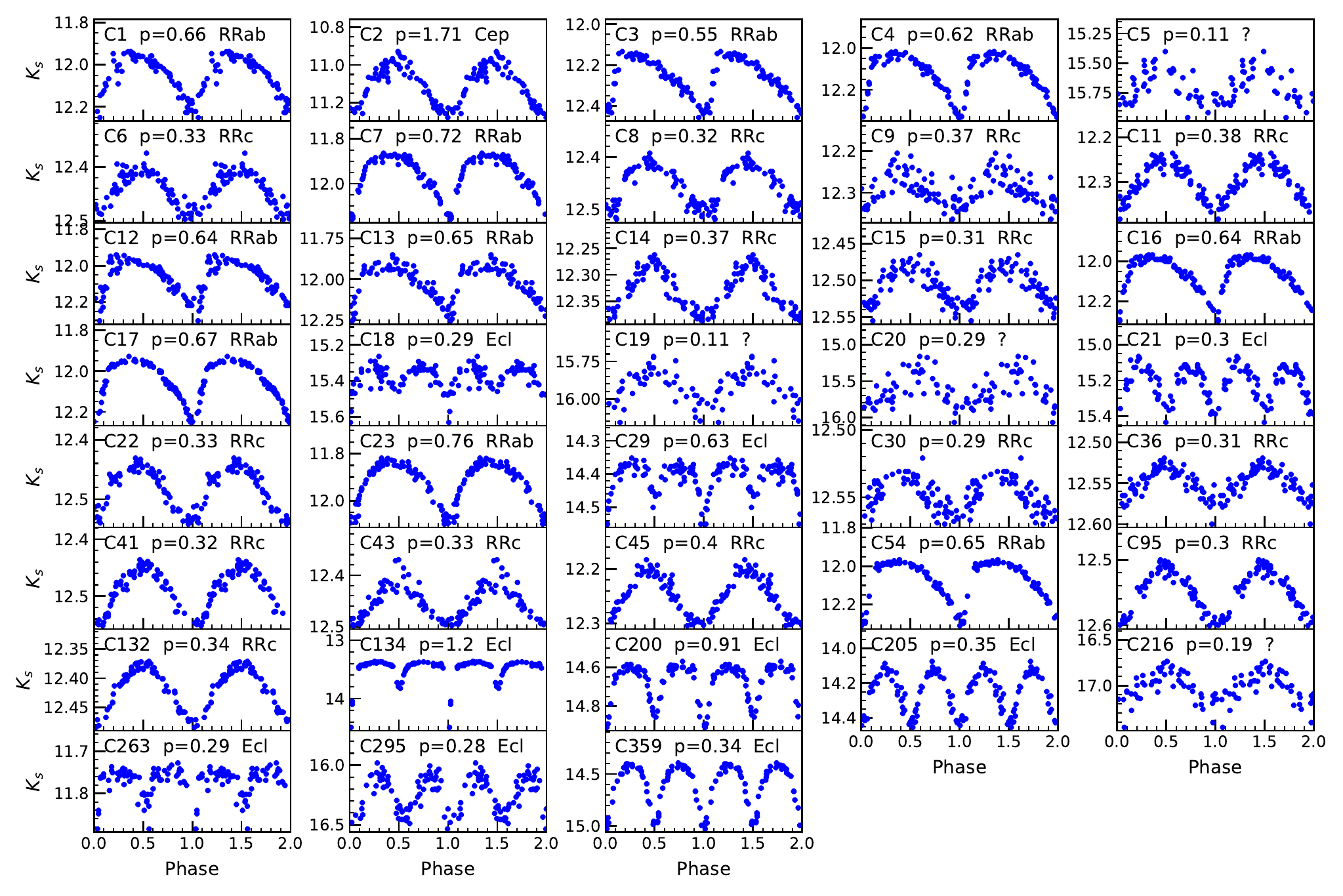}
  \caption{VVV $K_s$ phase-folded light curves for variable candidates
    in the M~22 field selected as cluster members, as shown in
    Sect.~\ref{sec_m22pm}. For every variable candidate, we provide
    its identifier, its (rounded) period in days, and its variable
    type (where available). The data for the light curves of all
      the variable candidates found in the studied area, including
      those used to create this figure, are available in electronic
      form at the CDS.}
  \label{fig_m22var}
\end{figure*}

There are 38 variable sources (including 10 RRab, 12 RRc, and 1
Cepheid) in our catalog with PMs consistent with being cluster members
according to our kNN classification criterion (see lower left and
right panels of Fig.~\ref{fig_m22}), which are highlighted in the
second to last column from Table~\ref{tab_m22var} and whose light
curves are shown in Fig.~\ref{fig_m22var}. Among these, we note 6
previously unreported variable sources: C5, C19, C20, and C216 are
variable stars with a short period and a sinusoidal, low-amplitude
light curve, which do not allow us to properly identify their variable
type; C263 and C359 are short period variables as well, but with
distinct eclipses, which lead us to classify them as eclipsing
binaries. However, the PMs of C5 and C20 are relatively far from the
mean of the cluster, which casts some doubts on their membership to
M~22. As expected, we observe in Table~\ref{tab_m22var} that most of
the innermost variables belong to the cluster according to our
classification criteria, but we also identify some variable members
relatively far from the cluster center. If confirmed as cluster
members by follow-up identification of their radial velocities, C275,
C307 and C374 would be the variables in M~22 farthest away from its
center.

In the right panel of Fig.~\ref{fig_m22} we present the near-infrared
CMD for M~22. The evolutionary sequences of this GGC can be clearly
observed to the blue of the red-giant branch (RGB) of the bulge field
stars. The cluster sequences stand out even more if we focus just on
the innermost ($r\leq 1'$) stars further selected according to our kNN
classifier. The upper main sequence (MS), RGB, and horizontal branch
(HB) of the cluster are clearly defined. The position in the CMD of
the RR~Lyrae selected using our membership criterion, clumped along
the HB, agrees with their expected position if they were cluster
members, reinforcing their high probabilities to belong to M~22.

\subsection{Distance and extinction}
\label{sec_m22dis}
Although such parameters as the extinction and the distance to M~22
could seem straightforward to calculate thanks to the tight
period-luminosity-metallicity (PLZ) relations that RR~Lyrae show in
the near-infrared \citep{lon86,cat04}, some assumptions need to be
made before proceeding to their calculation: 1) We report the
magnitudes of the detected variables in the VISTA near-infrared
photometric system \citep{gonfer18}. There are at least three main PLZ
relations for RR~Lyrae in this system: those by \citet{cat04}
calibrated in the VISTA system that we show in \citetalias{alo15};
those calibrated by \citet{mur15}; and those calibrated by
\citet{nav17,nav17b}. We checked the results each of these provide and
explored the required adjustments for each of these relations to
provide consistent results. 2) The PLZ relations mentioned in the
previous point are calibrated for their use with RRab stars. To use
the RRc sources as well, we need to fundamentalize their periods via
the relation $\log~P_{\rm ab}=\log~P_{\rm c}+0.127$
\citep[e.g.,][]{del06,nav17}.  3) We assume the metallicity of all the
stars in the cluster to be the same. Using the iron content of M~22
from the Harris catalog ($[{\rm Fe/H}]=-1.70$, see
Table~\ref{tab_gcinput}) and the ${\alpha}$-enhancement of
$[\alpha/{\rm Fe}]=0.33$ for M~22 by \citet{mar11a}, we obtained from
Eq. (1) in \citet{nav17} a value of $Z=0.0006$, assuming
$Z_{\odot}=0.017$ for consistency with \citet{cat04}. We note however
that most recent studies based on high-resolution spectroscopy or
narrow and medium-band photometry suggest M~22 to be one of the few
GGCs that show a spread in the iron-peak elements ($\sim0.1$ dex) in
addition to variations in the lighter elements
\citep{dac09,mar09,mar11,lee15,lee16}.  There is still some
controversy, however, with groups suggesting the spread in iron is not
real \citep{muc15} and others not being able to make a strong
statement on this matter \citep{mes20}. Such a spread, if real, would
not significantly alter our results. 4) We have observations for
filters $Z$, $Y$, $J$, and $H$ only in approximately two epochs per
filter. We assume that the mean $\lambda-K_s$ apparent colors obtained
from these few measurements following the method described in
Sect.~\ref{sec_var} is equal to the apparent colors obtained from the
mean magnitudes of the light curves. Since the observations in filters
$ZYJH$ were done at random phases for the different RR~Lyrae, the main
effect from this assumption in our estimation of the cluster
extinction is a slightly higher dispersion in the mean distribution of
the color excesses for the different filters, as we mentioned in
\citetalias{alo15}. For $J$ and $H$, we compared our results with
those obtained following the method described in \citet{haj18} to
calculate the magnitudes for the RR~Lyrae on those filters using only
a very limited number of epochs; we found no significant differences
on the mean extinction measured for the cluster.  5) Differential
extinction in M~22 \citep{alo12} is significantly smaller than in the
GGCs in \citetalias{alo15}, and therefore we are not able to find the
true distance and the selective-to-total extinction ratios $R$
simultaneously as we did there, and we need to assume some
selective-to-total extinction ratio to proceed with the distance
calculation. Extinction toward the inner Galaxy has been shown to be
non-canonical and highly variable \citep{alo17}, with several
measurements available in the literature
\citep[e.g.,][]{maj16,alo17,nog18,dek19}. However, given that M~22 is
not located at very low Galactic latitudes, we decided to use the
canonical law provided by \citet{car89}, which is good for the outer
halo of the Galaxy. Nevertheless, we also examined the effect of using
the extinction ratios provided by \citet{alo17}, which are good for
the innermost Galaxy, on the estimation of the distance to the
cluster.

\begin{table*}
  \caption{Color excesses and distances to the target clusters.}
  \label{tab_gcdis}
  \centering
  \resizebox{\textwidth}{!}{
  \begin{tabular}{cccccccc}
    \hline\hline
    Cluster & E($Z-K_s$) & E($Y-K_s$) & E($J-K_s$) & E($H-K_s$) & $R_{\odot}$ & $R_{\odot,\rm Baumgardt19}$ & $R_{\odot,\rm Harris96}$\\
     &(mag) & (mag) & (mag) & (mag) & (kpc) & (kpc) & (kpc)\\
    \hline
    NGC~6441 & 0.41$\pm$0.10 & 0.20$\pm$0.08 & 0.18$\pm$0.06 & 0.03$\pm$0.02 & 13.0$\pm$0.3+0.2+0.8 & 11.83$\pm$0.05 & 11.6\\
    Terzan~10 & 2.44$\pm$0.36 & 1.54$\pm$0.26 & 0.86$\pm$0.12 & 0.30$\pm$0.04 & 10.3$\pm$0.3-1.1+0.1 & -- & 5.8\\
    2MASS-GC~02 & -- & -- & 3.0$\pm$0.4 & 1.12$\pm$0.23 & 6.0$\pm$0.9 & -- & 4.9\\
    NGC~6569 & 0.65$\pm$0.05 & 0.37$\pm$0.04 & 0.25$\pm$0.05 & 0.07$\pm$0.03 & 10.6$\pm$0.3+0.3+0.3 & 10.24$\pm$1.16 & 10.9\\
    NGC~6626 (M~28) & 0.46$\pm$0.09 & 0.28$\pm$0.07 & 0.23$\pm$0.03 & 0.05$\pm$0.02 & 5.41$\pm$0.08+0.11-0.11 & 5.42$\pm$0.33 & 5.5\\
    NGC~6656 (M~22) & 0.35$\pm$0.07 & 0.24$\pm$0.06 & 0.18$\pm$0.04 & 0.05$\pm$0.03 & 3.24$\pm$0.05+0.06-0.04 & 3.24$\pm$0.08 & 3.2\\
    \hline
  \end{tabular}}
\tablefoot{The reported first $\sigma$ in both distances and color
  excesses corresponds to the dispersion from the individual RR~Lyrae
  estimations, the second $\sigma$ in the distance estimation
  corresponds to the effects of changing the extinction law that we
  use in the different GGCs according to the text, and the third to
  changing the PLZ relations.}
\end{table*}

\begin{figure}
  \centering
  \includegraphics[scale=0.65]{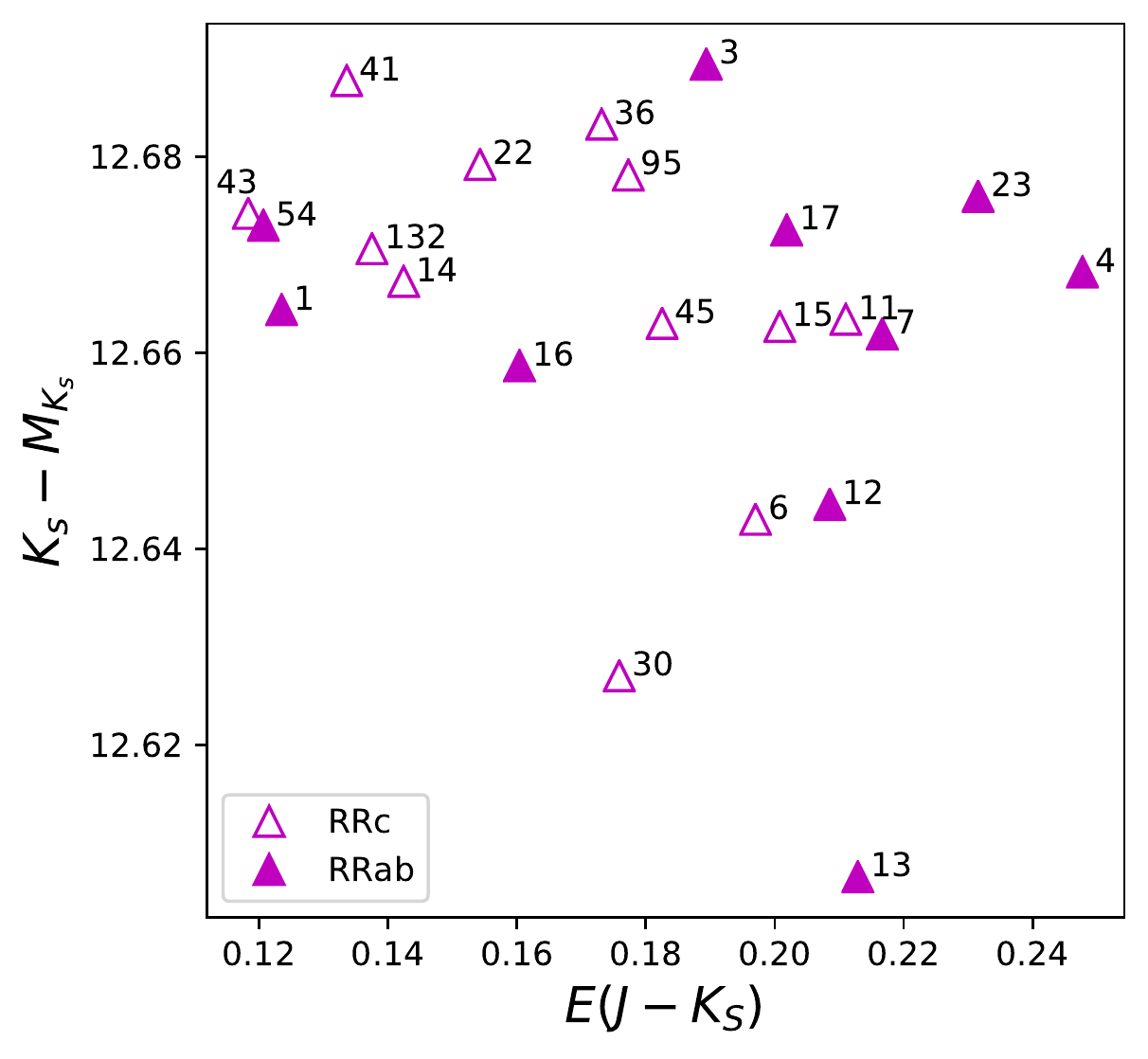}
  \caption{Distance modulus vs. $E(J-K_s)$ for those RR~Lyrae selected
    as members of M~22. The solid magenta triangles indicate RRab
    stars, while empty magenta triangles represent RRc stars. The
    distance moduli are not corrected for extinction. The relatively
    small near-infrared differential extinction baseline does not
    allow for a proper determination or selection of a particular
    extinction law.}
  \label{fig_m22dis}
\end{figure}

Taking the above points into consideration, our first calculations
using the sample of 22 RRLyrae that belong to M~22 (see
Sect.~\ref{sec_m22pm}) showed a significant variation between the
distance moduli based on the different PLZ relations we used:
$\mu_{K_s}=12.82\pm0.02$ mag from \citetalias{alo15},
$\mu_{K_s}=12.78\pm0.02$ mag from \citet{nav17b}, and
$\mu_{K_s}=12.64\pm0.02$ mag from \citet{mur15}. Using the PLZ
relations for the other VVV filters and the extinction ratios by
\citet{car89}, we were able to correct the distance moduli from
extinction. For \citet{mur15}, since there are no PLZ relations
provided in the other VVV filters that we could use to correct for
extinction, we took the color excesses provided by the PLZ relations
from \citetalias{alo15}. We obtained the following distances to M~22:
$3.48\pm0.05$ kpc from \citetalias{alo15}, $3.36\pm0.06$ kpc from
\citet{nav17b}, and $3.20\pm0.05$ kpc from \citet{mur15}. Only the
distance estimate provided by \citet{mur15} agrees with the values
provided in the literature (see Table~\ref{tab_gcdis}). We note that
the PLZ relations by \citet{nav17b} were calibrated using a distance
modulus for $\omega$ Centauri, $\mu_0=13.708$ mag, a value somewhat
higher than the latest measurements from {\em Gaia}, $\mu_0=13.60$ mag
\citep{bau19}. If we calibrate the PLZ relations from \citet{nav17b}
using the latter $\mu_0$ value for $\omega$ Centauri, we need to apply
an offset of 0.11 to Eqs. (4) and (5) from \citet{nav17b}, leaving
them as
   \begin{equation}
  M_J(RRL)=-1.77 \log(P) + 0.15 [{\rm Fe/H}] - 0.45,
   \end{equation}
   \begin{equation}
  M_{K_s}(RRL)=-2.23 \log(P) + 0.14 [{\rm Fe/H}] - 0.78.
   \end{equation}
Applying these updated PLZ relations, we obtained for M~22 a distance
modulus $\mu_{K_s}=12.67\pm0.02$ mag, and after correction for
extinction, a distance of $3.20\pm0.06$ kpc, which better agrees with
values from the literature (see Table~\ref{tab_gcdis}). As the PLZ
relations from \citetalias{alo15} seem to provide consistent results
according to the examination done in \citet{nav17}, we assumed that
there was only an offset in their calibration. Assuming the distance to
M~22 to be 3.24 kpc from the {\em Gaia} determination by
\citet{bau19}, we found the offset we need to apply to the PLZ
relations from \citetalias{alo15} to be 0.157, transforming them into
   \begin{equation}
     M_{K_s} = -0.480 - 2.347 \log(P) + 0.1747 \log(Z),
   \end{equation}
   \begin{equation}
     M_H  = -0.397 - 2.302 \log(P) + 0.1781 \log(Z),
   \end{equation}
   \begin{equation}
     M_J  = -0.079 - 1.830 \log(P) + 0.1886 \log(Z),
   \end{equation}
   \begin{equation}
     M_Y  = +0.166 - 1.467 \log(P) + 0.1966 \log(Z),
   \end{equation}
   \begin{equation}
     M_Z  = +0.314 - 1.247 \log(P) + 0.2014 \log(Z).
   \end{equation}
Using these updated PLZ relations from \citetalias{alo15}, we show in
Fig.~\ref{fig_m22dis} the distance moduli and color excesses for all
22 RR~Lyrae selected in Sect.~\ref{sec_m22pm} as M~22 members. We
stress the small dispersions they show.  We report the mean values of
the distance and color excesses for M~22 in Table~\ref{tab_gcdis}.

In principle, we could also use C2 (V24), the Cepheid we detected, to
calculate distance and extinction to M~22. Unfortunately, the PL
relations currently calculated for Type~2 Cepheids in the VISTA
photometric system only cover the $J$, $H$, and $K_s$ filters
\citep{bha17,bra19,dek19}. C2 is saturated in $J$ and $H$ (see
Table~\ref{tab_m22var}), so we were not able to obtain the extinction
from the PL relations. Assuming for C2 the mean extinction for the
cluster reported in Table~\ref{tab_gcdis} from the RR~Lyrae estimation
and the \citet{car89} extinction law, we obtained a distance of
$3.46\pm0.02$ kpc. This distance is a little higher than the value we
obtained from the literature (see Table~\ref{tab_gcdis}), but
interestingly it agrees well with the values we were obtaining from
the original PLZ relations from \citetalias{alo15} before we
recalibrated them. However, we note that our sample of Cepheids in
this GGC consists of only one detected star, and therefore all results
obtained from the analysis of such a small sample should be regarded
with caution.

\section{NGC~6626 (M~28), NGC~6569, and NGC~6441}
\label{sec_gc1}
NGC~6441, NGC~6569, and NGC~6626 (M~28) are also massive GGCs located
toward the inner Galaxy (see Table~\ref{tab_gcinput}), although, as
seen in the reddening maps of these regions \citep{gon12,gon18}, they
are at Galactic latitudes where the values of extinction are still not
as extreme as for the GGCs in Sect.~\ref{sec_gc2}. As shown in
Table~\ref{tab_gcinput}, the metallicities of these GGCs extend over a
wide range, and, as shown in Table~\ref{tab_gcdis}, their kinematic
distances were recently measured using {\em Gaia} DR2 \citep{bau19}.

\subsection{Variables in the cluster area}
\label{sec_gc1var}
A significant number of variable stars, including several classical
pulsators, are already known in the regions covered by these
GGCs. However, as for M~22 in Sect.~\ref{sec_m22}, ours is the first
study to characterize the populations of variable stars in these GGCs
in the near-infrared, covering the whole cluster area.

M~28 was classified as an Oosterhoff intermediate or a hybrid
Oosterhoff~I/II system by \citet{pri12}, the most recent study of the
variable stars in this GGC. We detected and characterized 88 variable
candidates in the field of M~28. We present their main observational
parameters in Table~\ref{tab_m28var}.  From the 13(18) RRab stars
reported in the Clement(OGLE) catalog, we only fail to detect 1(2). As
expected, the proportion of detected RRc sources is a little
lower. From 9(10) RRc stars reported in the Clement(OGLE) catalog, we
detected 3(5) of them. Furthermore, there are 2 reported Cepheids in
the Clement catalog, but they appear neither in the OGLE catalog nor
in ours, probably due to saturation. Finally, we detected 66 new
variable candidates. We classify 2 as RRab, 13 as eclipsing binaries,
and we are not able to assign a variable type to the other 51
candidates.

\begin{table*}
  \caption{Properties of the variable candidates in NGC~6626 (M~28).}
  \label{tab_m28var}
  \centering
  \resizebox{\textwidth}{!}{
  \begin{tabular}{ccccccccccccccccc}
    \hline\hline
    ID & $ID_{Clement}$ & $ID_{OGLE}$ & $\alpha$ (J2000) & ${\delta}$ (J2000) & Distance\tablefootmark{a} & Period & $A_{K_s}$ & $\langle K_s \rangle$ & $Z-K_s$ & $Y-K_s$ & $J-K_s$ & $H-K_s$ & $\mu_{\alpha^\ast}$ & $\mu_{\delta}$ & Member\tablefootmark{b} & Type \\
    & & (OGLE-BLG-) & (h:m:s) & (d:m:s) & (arcmin) & (days) & (mag) & (mag)& (mag)& (mag)& (mag) & (mag) & (mas/yr) & (mas/yr) & & \\ 
    \hline
C1 & V20 & RRLYR-59819 & 18:24:33.11 & -24:51:43.0 & 0.48 & 0.497742 & 0.262 & 13.459 & 0.689 & 0.525 & 0.407 & -0.002 & -4.975 & -7.654 & Yes & RRab\\
C2 & V22 & RRLYR-59799 & 18:24:30.98 & -24:52:01.9 & 0.49 & 0.322637 & 0.107 & 13.637 & 0.738 & 0.448 & 0.366 & 0.088 & 1.108 & -12.76 & Yes & RRc\\
C3 & V23 & RRLYR-59795 & 18:24:30.25 & -24:52:03.1 & 0.64 & 0.292314 & 0.09 & 13.774 & 0.766 & 0.528 & 0.367 & 0.12 & 3.496 & -8.488 & Yes & RRc\\
C4 & V11 & RRLYR-59804 & 18:24:31.48 & -24:51:33.1 & 0.73 & 0.54276 & 0.329 & 13.41 & 0.889 & 0.61 & 0.426 & 0.118 & -1.411 & -9.161 & Yes & RRab\\
C5 & V18 & RRLYR-59844 & 18:24:36.54 & -24:51:50.1 & 0.88 & 0.640177 & 0.293 & 13.252 & 1.068 & 0.787 & 0.504 & 0.119 & 4.428 & -7.115 & Yes & RRab\\
C6 & V25 & RRLYR-59787 & 18:24:28.82 & -24:52:09.1 & 0.95 & 0.74772 & 0.206 & 13.105 & 1.0 & 0.731 & 0.579 & 0.099 & 1.732 & -7.355 & Yes & RRab\\
C7 & V29 & RRLYR-59780 & 18:24:27.66 & -24:52:15.1 & 1.21 & 0.311162 & 0.085 & 13.6 & 0.882 & 0.611 & 0.444 & 0.103 & -0.551 & -8.168 & Yes & RRc\\
C8 & V13 & RRLYR-59773 & 18:24:25.77 & -24:52:33.7 & 1.68 & 0.654924 & 0.294 & 13.247 & 1.067 & 0.782 & 0.504 & 0.131 & -1.095 & -8.659 & Yes & RRab\\
C9 & -- & -- & 18:24:24.61 & -24:53:01.8 & 2.08 & 0.172628 & 0.479 & 14.657 & 1.192 & 0.862 & 0.622 & 0.147 & 1.698 & -5.384 & No & ?\\
C10 & V12 & RRLYR-59878 & 18:24:43.31 & -24:52:56.3 & 2.45 & 0.578212 & 0.333 & 13.367 & 0.943 & 0.676 & 0.51 & 0.121 & 0.262 & -8.341 & Yes & RRab\\
    \hline
  \end{tabular}}
\tablefoot{This table is available in its entirety in electronic form
  at the CDS.  A portion is shown here for guidance regarding its form and content.\\
\tablefoottext{a}{Projected distance to the cluster center}
\tablefoottext{b}{Cluster membership according to criteria explained in Sect.~\ref{sec_gc1dis}}
}
\end{table*}

NGC~6569 could be considered an Oosterhoff~I GGC according to its
metallicity and mean periods of their RRab, although the high ratio of
RRc and RRab, characteristic of an Oosterhoff~II GGC, casts some
doubts on its proper Oosterhoff classification \citep{kun15}. We
detected 27 variable candidates in the field of NGC 6569. We present
the main observational parameters of these candidates in
Table~\ref{tab_ngc6569var}. Detection of the variable stars seems
highly dependent on their distance to the cluster center. While at
distances $1.1' \leq r \leq r_t$ we recovered all the 4(6) RRab stars
from the Clement(OGLE) catalog, at distances $r<1.1'$ from the cluster
center we only recovered 1(1) of the 9(6) RRab sources shown in the
Clement(OGLE) catalog. As expected, the proportion of detected RRc
stars is lower. From the 12(13) RRc stars reported in the
Clement(OGLE) catalog, we failed to detect all 4(4) stars at distances
$r<1.1'$ from the cluster center, and from the remaining ones at $1.1'
\leq r \leq r_t$, we only detected 2(2) of them. Furthermore, there is
1 reported Cepheid in the Clement catalog, which saturates in our
photometry. There is another one in the OGLE catalog that we were not
able to recover either. Finally, we detected 6 new variable candidates
in NGC~6569. We classified C2 as an RRab, C19 as a Cepheid, and we
were not able to assign a variable type to the other 4 candidates.

\begin{table*}
  \caption{Properties of the variable candidates in NGC~6569.}
  \label{tab_ngc6569var}
  \centering
  \resizebox{\textwidth}{!}{
  \begin{tabular}{ccccccccccccccccc}
    \hline\hline
    ID & $ID_{Clement}$ & $ID_{OGLE}$ & $\alpha$ (J2000) & ${\delta}$ (J2000) & Distance\tablefootmark{a} & Period & $A_{K_s}$ & $\langle K_s \rangle$ & $Z-K_s$ & $Y-K_s$ & $J-K_s$ & $H-K_s$ & $\mu_{\alpha^\ast}$ & $\mu_{\delta}$ & Member\tablefootmark{b} & Type \\
    & & (OGLE-BLG-) & (h:m:s) & (d:m:s) & (arcmin) & (days) & (mag) & (mag)& (mag)& (mag)& (mag) & (mag) & (mas/yr) & (mas/yr) & & \\ 
    \hline
C1 & V26 & RRLYR-34968 & 18:13:36.88 & -31:49:17.2 & 0.54 & 0.653784 & 0.295 & 14.761 & 1.196 & 0.813 & 0.465 & 0.132 & -3.418 & -7.771 & Yes & RRab\\
C2 & -- & -- & 18:13:35.08 & -31:50:15.6 & 1.07 & 0.672413 & 0.212 & 14.898 & 1.223 & 0.855 & 0.51 & 0.128 & -3.371 & -7.906 & Yes & RRab\\
C3 & V17 & RRLYR-34982 & 18:13:39.21 & -31:50:46.5 & 1.19 & 0.53611 & 0.33 & 15.087 & 1.048 & 0.684 & 0.522 & 0.103 & -5.731 & -5.845 & Yes & RRab\\
C4 & V2 & RRLYR-34949 & 18:13:31.04 & -31:49:40.7 & 1.69 & 0.574729 & 0.286 & 15.004 & 1.073 & 0.74 & 0.516 & 0.122 & -2.682 & -5.495 & Yes & RRab\\
C5 & V37 & ECL-369682 & 18:13:33.21 & -31:50:58.5 & 1.86 & 0.631562 & 0.19 & 15.166 & 0.996 & 0.667 & 0.473 & 0.128 & -0.234 & -2.314 & No & Ecl\\
C6 & V38 & ECL-370166 & 18:13:37.57 & -31:47:31.6 & 2.08 & 0.947012 & 0.196 & 15.059 & 1.103 & 0.758 & 0.483 & 0.148 & -6.705 & -3.783 & No & Ecl\\
C7 & V20 & RRLYR-35010 & 18:13:45.97 & -31:51:13.5 & 2.21 & 0.542082 & 0.357 & 15.011 & 1.187 & 0.807 & 0.44 & 0.135 & -4.884 & -8.345 & Yes & RRab\\
C8 & -- & -- & 18:13:30.70 & -31:51:04.0 & 2.3 & 0.138767 & 0.455 & 14.685 & 1.508 & 1.079 & 0.731 & 0.187 & 3.751 & -6.034 & No & ?\\
C9 & -- & -- & 18:13:33.94 & -31:47:30.8 & 2.33 & 0.210626 & 0.081 & 11.783 & 0.637 & 0.429 & 0.339 & 0.101 & -3.608 & 0.242 & No & ?\\
C10 & V12 & RRLYR-34936 & 18:13:27.37 & -31:49:56.1 & 2.49 & 0.261068 & 0.093 & 15.351 & 0.827 & 0.533 & 0.33 & 0.092 & -2.724 & -8.771 & Yes & RRc\\
    \hline
  \end{tabular}}
\tablefoot{This table is available in its entirety in electronic form
  at the CDS. A portion is shown here for guidance regarding its form and content.\\
\tablefoottext{a}{Projected distance to the cluster center}
\tablefoottext{b}{Cluster membership according to criteria explained in Sect.~\ref{sec_gc1dis}}
}
\end{table*}

Lastly, NGC~6441 is one of the most intriguing GGCs in the Milky Way
on account of its radial pulsators. Along with NGC~6388 \citep{pri02},
NGC~6441 belongs to the Oosterhoff~III GGCs, which is characterized
for being metal-rich and hosting long-period RR~Lyrae. In this regard,
NGC~6441 has been the subject of several studies to look at its
population of variable stars
\citep[e.g.,][]{pri01,pri03,cor06,kun18}. We identified 59 variable
candidates in our analysis of the region covered by this GGC. We
present their main observational parameters in
Table~\ref{tab_ngc6441var}. As for NGC~6569, our detection of the
variable stars depends highly on their distance to the cluster
center. While from the 20(20) RRab sources from the Clement(OGLE)
catalog at distances $1.0'\leq r \leq r_t$, we missed only 4(3), at
distances $r<1.0'$ from the cluster center we only recovered 2(2) of
the 31(12) RRab stars shown in the Clement(OGLE) catalog. As expected,
the detection of RRc stars with good-quality light curves is worse. We
did not detect any of the 15(5) RRc sources reported in the
Clement(OGLE) catalog at distances $r<1'$, while we only recovered
4(3) of the 13(11) RRc stars reported in the Clement(OGLE) catalog at
distances $1.0'\leq r \leq r_t$. Moreover, we recovered none of the
8(2) Cepheids reported in the Clement(OGLE) catalog at distances
$r<1.0'$, but we recovered the only Cepheid reported in the OGLE
catalog at distances $1.0'\leq r \leq r_t$. Furthermore, we report 9
new variable candidates. Unfortunately, we were not able to assign a
variable type to any of these. Finally, it is worth highlighting 3 of
the variable stars present in NGC~6441: C8 (V150), C17 (V45), and C24
(V69).  For C8, we found a period much longer than reported in the
Clement catalog. A period this long ($\sim1.06$ days) for a pulsator
in a canonical GGC means C8 is a Cepheid. But as shown in
Sect.~\ref{sec_gc1dis}, C8 seems to follow the PLZ relations for an
RRab star in NGC~6441, so we kept its classification as an RRab. C24
is classified as an RRc in the Clement catalog and as an RRab in the
OGLE catalog. Even though its period ($\sim0.56$ days) suggests this
pulsator to be an RRab, its low amplitude suggests it is an RRc. In
Sect.~\ref{sec_gc1dis} we show that in order for it to be at the same
distance as the rest of RR~Lyrae in NGC~6441, it needs to be
considered as an RRc. The case of C17 could be similar to that of
C24. Its position in the CMD coincides with the RR~Lyrae from
NGC~6441, but its period, much shorter than the mean periods of the
RRab stars in NGC~6441, may suggest we are dealing with a long-period
RRc from this GGC. However, given that it has a higher amplitude than
C24, and its belonging to NGC~6441 is just borderline according to our
kNN classifier (see Sect.~\ref{sec_gc1pm}), we prefer to consider it
as a field RRab, as suggested by \citet{pri01}.

\begin{table*}
  \caption{Properties of the variable candidates in NGC~6441.}
  \label{tab_ngc6441var}
  \centering
  \resizebox{\textwidth}{!}{
  \begin{tabular}{ccccccccccccccccc}
    \hline\hline
    ID & $ID_{Clement}$ & $ID_{OGLE}$ & $\alpha$ (J2000) & ${\delta}$ (J2000) & Distance\tablefootmark{a} & Period & $A_{K_s}$ & $\langle K_s \rangle$ & $Z-K_s$ & $Y-K_s$ & $J-K_s$ & $H-K_s$ & $\mu_{\alpha^\ast}$ & $\mu_{\delta}$ & Member\tablefootmark{b} & Type \\
    & & (OGLE-BLG-) & (h:m:s) & (d:m:s) & (arcmin) & (days) & (mag) & (mag)& (mag)& (mag)& (mag) & (mag) & (mas/yr) & (mas/yr) & & \\ 
    \hline
C1 & V57 & RRLYR-03918 & 17:50:10.40 & -37:02:55.9 & 0.54 & 0.694358 & 0.299 & 15.228 & 0.949 & 0.614 & 0.56 & 0.114 & -3.486 & -9.286 & Yes & RRab\\
C2 & V59 & RRLYR-03904 & 17:50:08.95 & -37:03:32.2 & 0.94 & 0.702823 & 0.416 & 15.347 & 0.861 & 0.582 & 0.451 & 0.042 & -7.74 & -0.528 & Yes & RRab\\
C3 & V66 & RRLYR-03934 & 17:50:11.45 & -37:02:07.0 & 1.0 & 0.860932 & 0.174 & 14.989 & 1.046 & 0.703 & 0.461 & 0.12 & -2.915 & -7.391 & Yes & RRab\\
C4 & V61 & RRLYR-03980 & 17:50:14.69 & -37:04:02.6 & 1.03 & 0.750108 & 0.278 & 15.173 & 1.031 & 0.706 & 0.467 & 0.092 & -1.965 & -4.983 & Yes & RRab\\
C5 & V62 & RRLYR-03961 & 17:50:13.19 & -37:04:06.5 & 1.04 & 0.679969 & 0.294 & 15.167 & 0.727 & 0.51 & 0.577 & 0.12 & -2.832 & -3.402 & Yes & RRab\\
C6 & V40 & RRLYR-03920 & 17:50:10.51 & -37:04:00.5 & 1.07 & 0.648005 & 0.343 & 15.3 & 1.024 & 0.705 & 0.389 & 0.087 & -0.95 & -5.497 & Yes & RRab\\
C7 & V43 & RRLYR-04003 & 17:50:16.85 & -37:02:09.7 & 1.19 & 0.773081 & 0.248 & 15.086 & 1.019 & 0.697 & 0.51 & 0.088 & -1.63 & -7.458 & Yes & RRab\\
C8 & V150 & -- & 17:50:07.04 & -37:03:15.9 & 1.21 & 1.068624 & 0.184 & 14.709 & 1.058 & 0.754 & 0.467 & 0.11 & -2.487 & -5.487 & Yes & RRab\\
C9 & -- & -- & 17:50:05.90 & -37:03:20.7 & 1.44 & 2.93268 & 0.201 & 14.648 & 1.396 & 1.021 & 0.704 & 0.169 & -2.486 & -6.663 & Yes & ?\\
C10 & V42 & RRLYR-03956 & 17:50:13.05 & -37:01:31.5 & 1.54 & 0.812634 & 0.239 & 15.078 & 1.143 & 0.804 & 0.618 & 0.233 & -2.412 & -5.718 & Yes & RRab\\
    \hline
  \end{tabular}}
\tablefoot{This table is available in its entirety in electronic form
  at the CDS. A portion is shown here for guidance regarding its form and content.\\
\tablefoottext{a}{Projected distance to the cluster center}
\tablefoottext{b}{Cluster membership according to criteria explained in Sect.~\ref{sec_gc1dis}}
}
\end{table*}

\subsection{Proper motions, color-magnitude diagrams, and cluster memberships}
\label{sec_gc1pm}
\begin{figure*}
  \centering
  \begin{tabular}{cc}
  \includegraphics[scale=0.37]{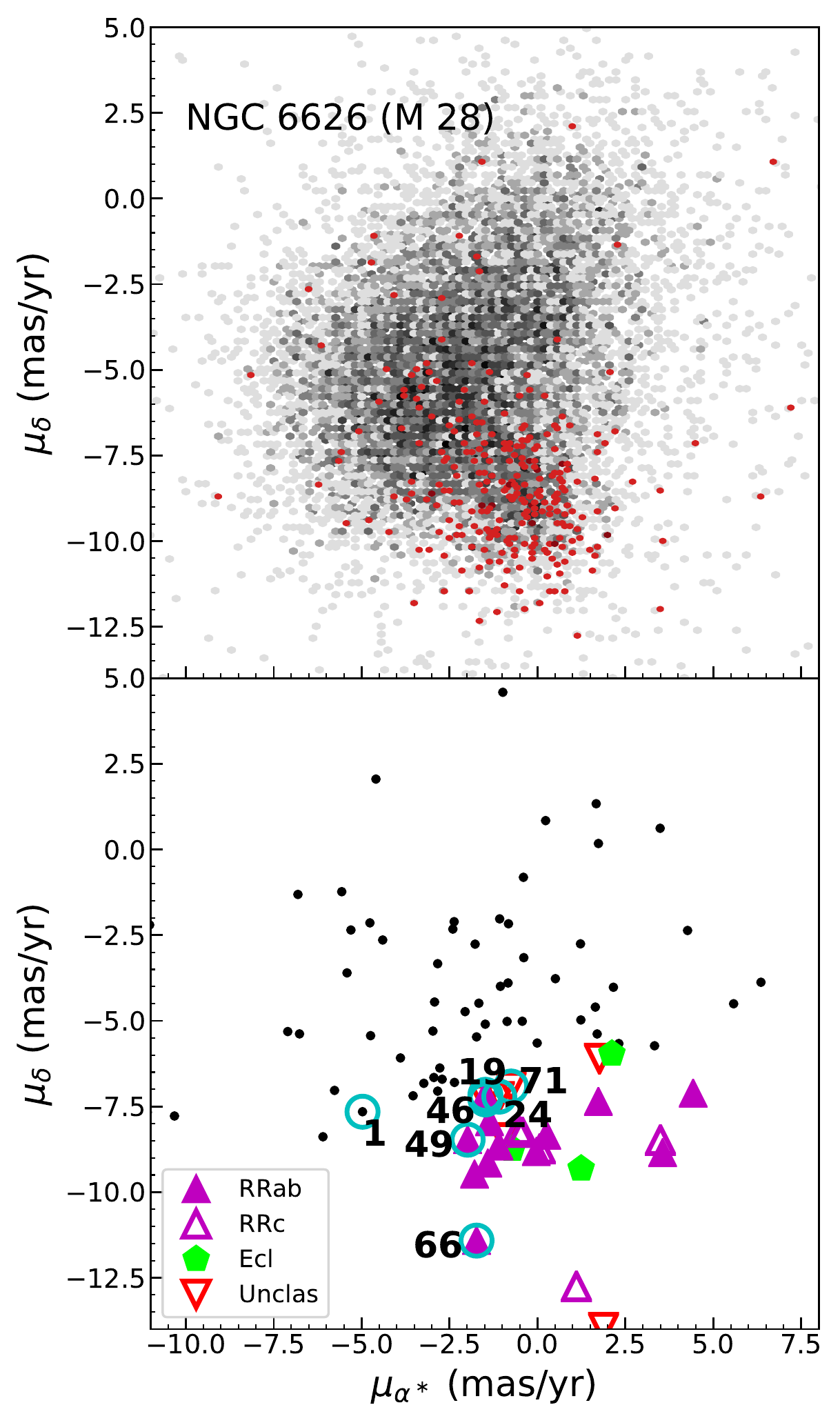}
  \includegraphics[scale=0.37]{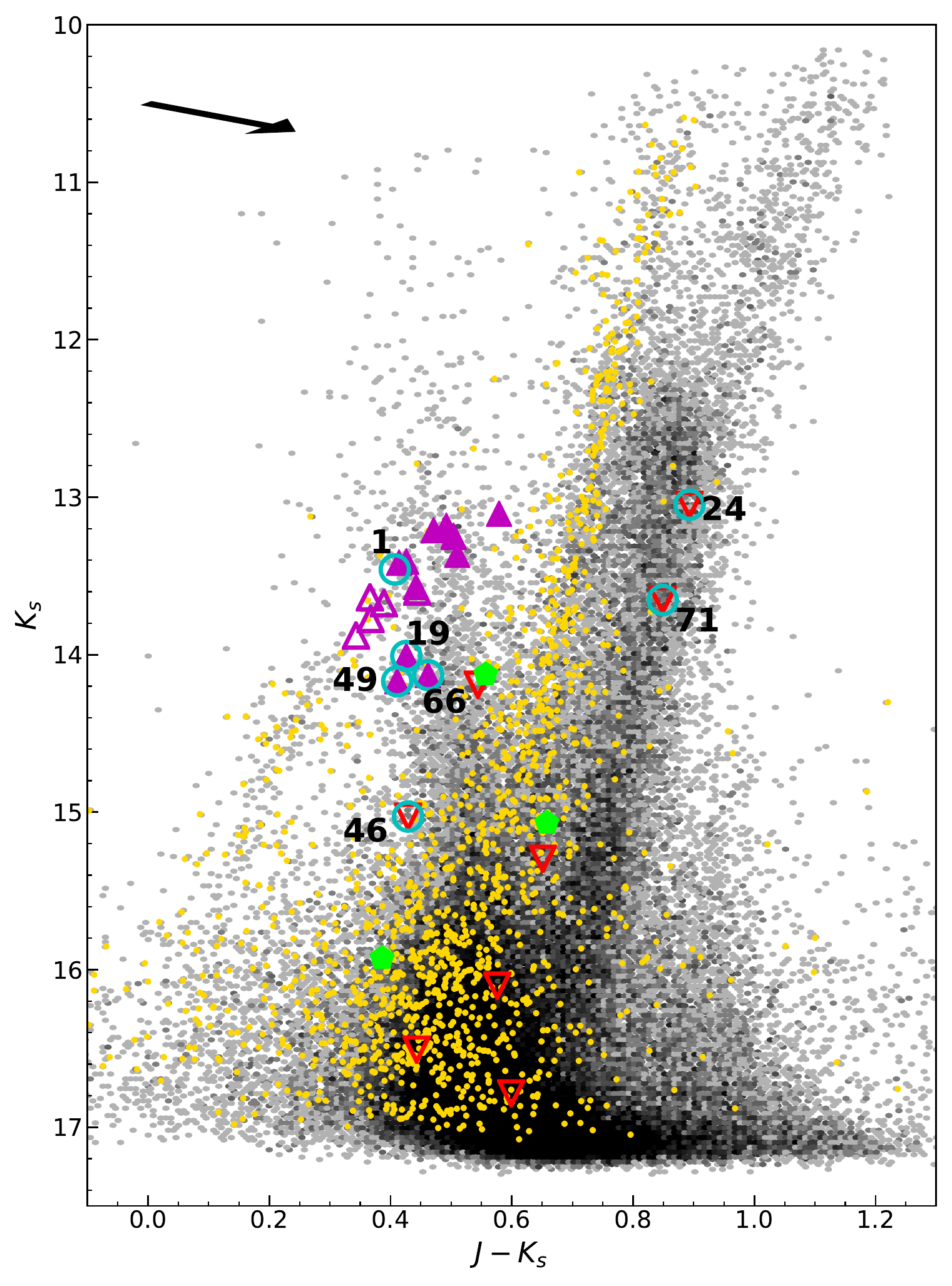}\\
  \includegraphics[scale=0.37]{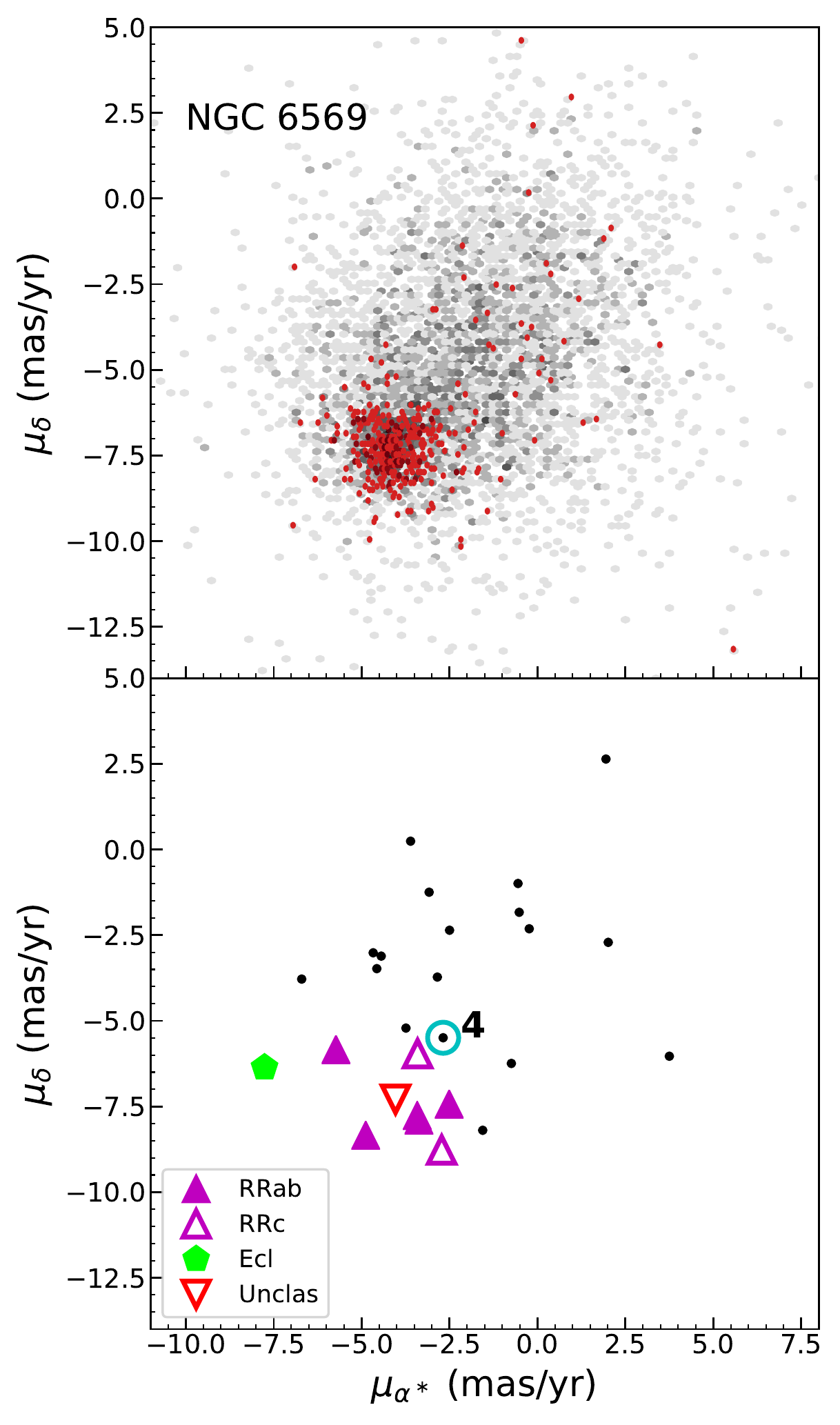}
  \includegraphics[scale=0.37]{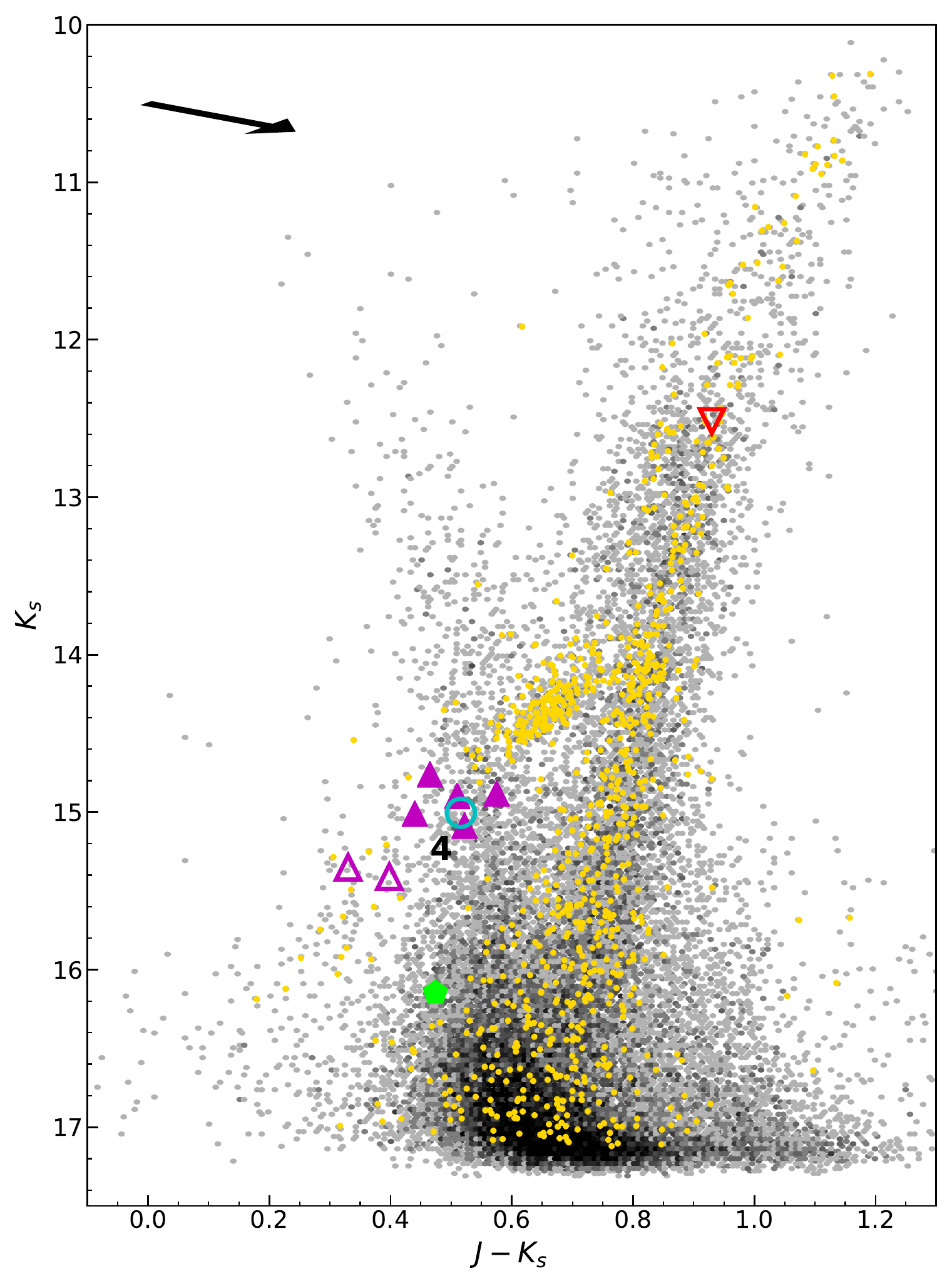}\\
  \includegraphics[scale=0.37]{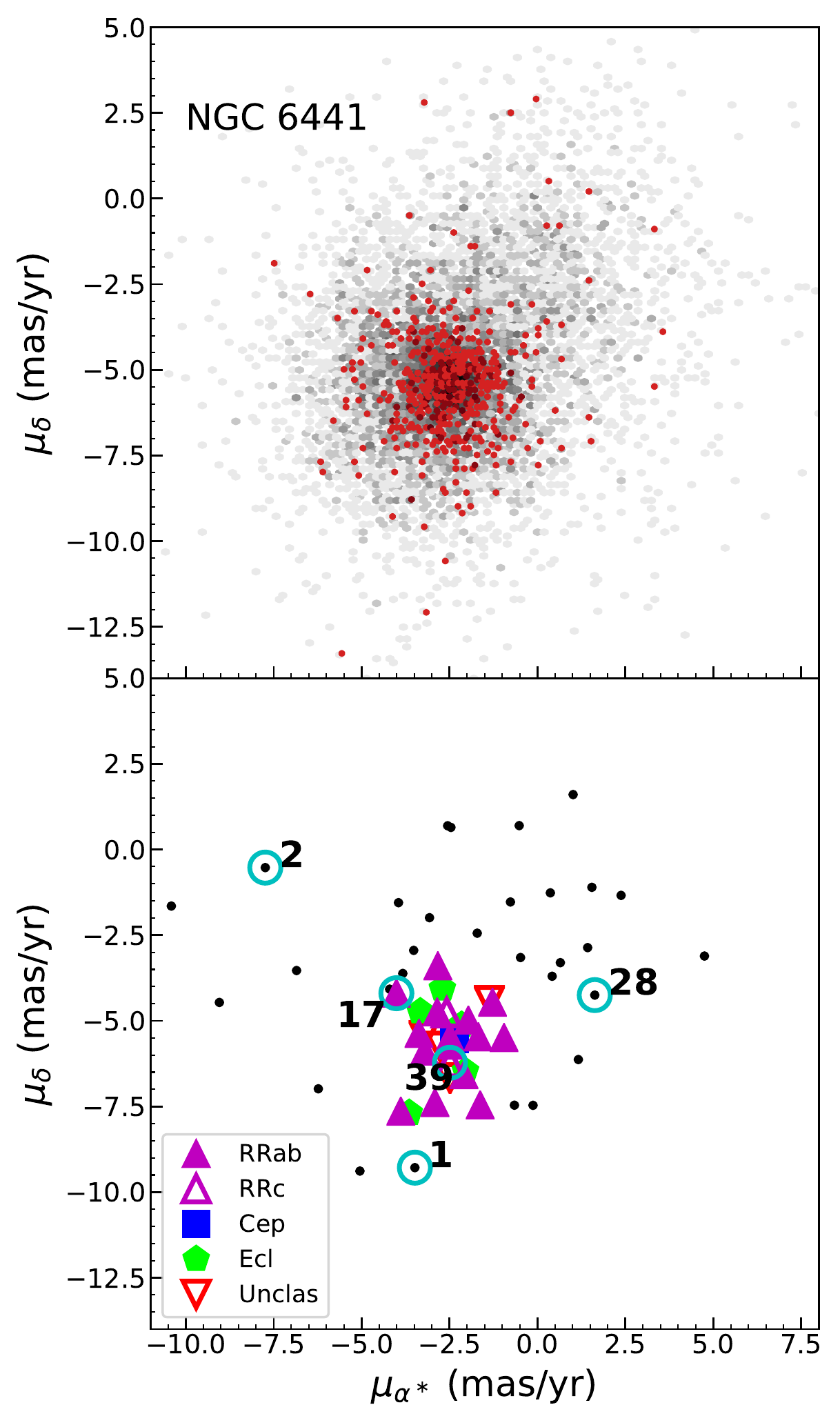}
  \includegraphics[scale=0.37]{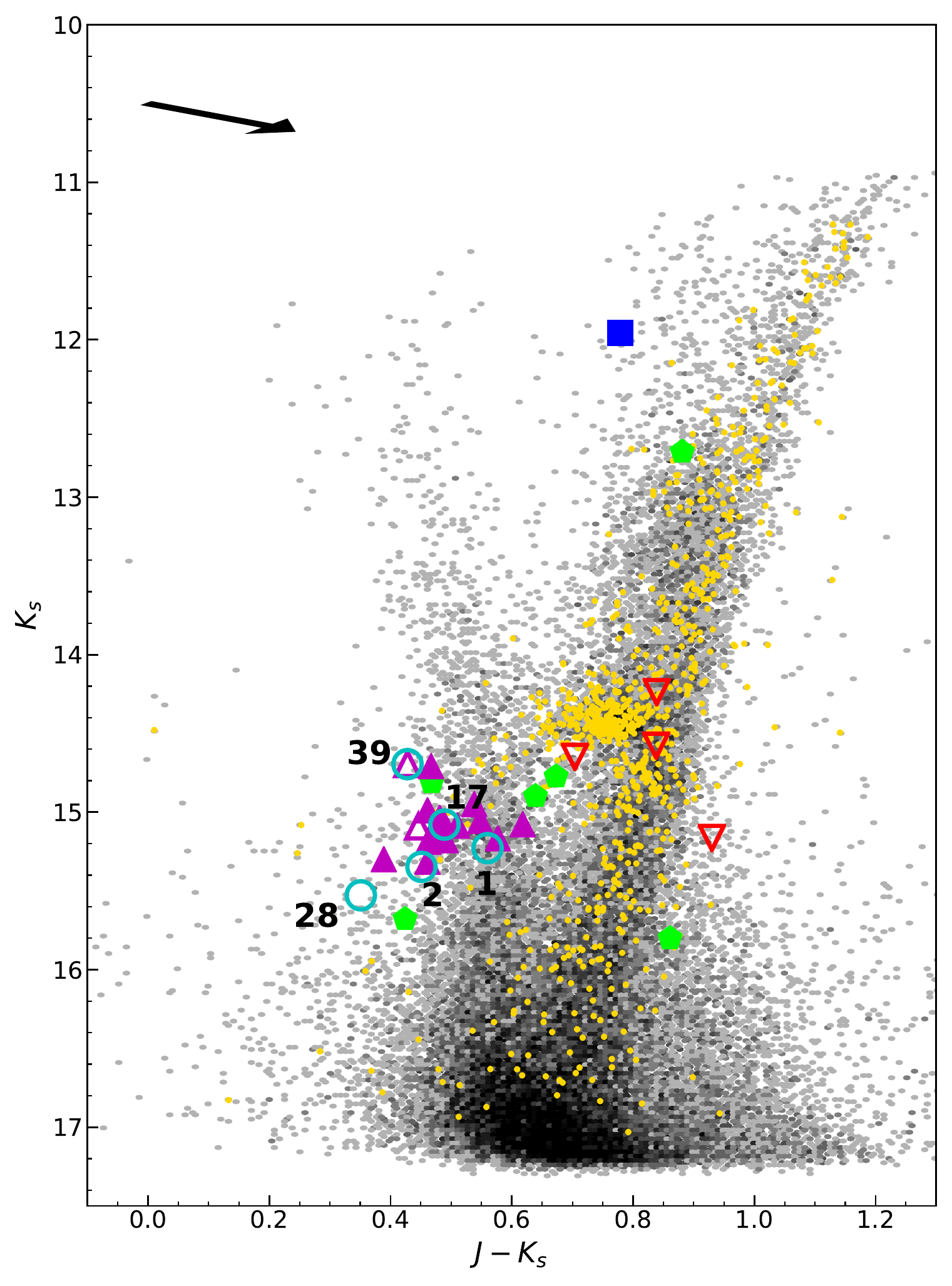}\\
  \end{tabular}
  \caption{As in Fig.~\ref{fig_m22}, but for NGC~6626 (M~28) in the
    top panels, for NGC~6569 in the central panels, and for NGC~6441
    in the lower panels. Now cyan empty circles encapsulate those
    variable candidates whose memberships to the cluster according to
    our kNN classifier were reversed (see text).}
  \label{fig_gc1}
\end{figure*}

We used the PMs provided by VIRAC2 for the detected variable stars to
assign membership to the GGCs. As we did for M~22 in
Sect.~\ref{sec_m22pm}, first we identified stars with precise PM
($\sigma_{\mu_{\alpha^\ast}}<1$ mas/yr, $\sigma_{\mu_{\delta}}<1$
mas/yr) and located at distances close to the cluster center
($r\le1'$) in these three GGCs (see left upper panels of
Fig.~\ref{fig_gc1}) to define the cluster membership criterion through
our kNN classifier. To train the classifier, we used the same criteria
that we used in M~22 (see Sect.~\ref{sec_m22pm}). While the separation
between the PMs of cluster and field populations is not as clear as for
M~22, in Table~\ref{tab_gcpm} we can see that the PMs of these GGCs,
defined by the mean of the PMs of the innermost ($r\le1.0'$) stars
selected by our kNN classifier, closely agree with those provided by
{\em Gaia} \citep{bau19}.

In the right panels of Fig.~\ref{fig_gc1} we present the CMDs for
these GGCs out to their tidal radii. Since they are contaminated by
field stars, we overplotted their innermost ($r\le1'$) stars selected
as members by our kNN classifier. We can appreciate that the dimmest
evolutionary sequences we reach in our CMDs for the inner regions of
M~28, the lower RGB and upper MS of the cluster appear well populated;
however, this is not the case for the inner regions of NGC~6569 and
NGC~6441, suggesting that for the innermost regions of these 2 GGCs
incompleteness for these evolutionary sequences is higher than for
M~28 and M~22. This would explain the lower number of detected
RR~Lyrae and Cepheids close to the cluster center shown in
Sect.~\ref{sec_gc1var} for both NGC~6569 and NGC~6441.

In the left lower panels of Fig.~\ref{fig_gc1} we present the PMs of
the variable candidates detected in the fields of the 3 GGCs
considered here, highlighting those that our kNN classifier selected
as cluster members. There are 28 member variable stars in M~28, 9 in
NGC~6569, and 28 in NGC~6441, according to our kNN classifier. When we
analyzed their position in the CMD, the bulk of variable candidates
seem to agree with their PM membership classification, but a few of
them needed their assigned memberships to be reversed. In the case of
M~28 (see top panels in Fig.~\ref{fig_gc1}), the positions in the CMD
of C24 and C71 (unclassified variables), and C19 (RRab) seem more
consistent with their classification as field stars. All of these
sources have similar PMs and are near the membership limit according to
the kNN classifier, which made us discard these as member stars, along
with C46 (unclassified), whose PM is also similar. We also discarded 2
other RRab stars, C49 and C66, owing to their positions in the
CMD. They are around half a magnitude below the other RRab sources and they
do not suffer from significant reddening. On the other hand, we
consider another RRab, C1, to be a member of M~28 owing to its
position in the CMD, even though the kNN classifier regards it as a
field star. We speculate that the projected proximity of this source
to the cluster center may influence the accuracy of its reported PM. A
similar situation seems to be happening for C1 and C2 in NGC~6441,
which we also classified as cluster members even though their PMs are
very different from those of the cluster members (see lower panels in
Fig.~\ref{fig_gc1}). The cases for the RR~Lyrae C28 in NGC~6441 and C4
in NGC~6569 are not so extreme. They have PMs closer to the bulk of the
stars in their respective GGCs. We classified these as as cluster
members according to their position in the CMD even though our kNN
classifier considers them to be field stars (see central and lower
panels of Fig.~\ref{fig_gc1}). Finally, we consider C39 in NGC~6441 to
be a field RRc star given its bright magnitude, even though the PM of
this source seems to suggest that it belongs to this GGC (see lower
panels in Fig.~\ref{fig_gc1}).

\begin{figure*}
  \centering
  \includegraphics[scale=0.85]{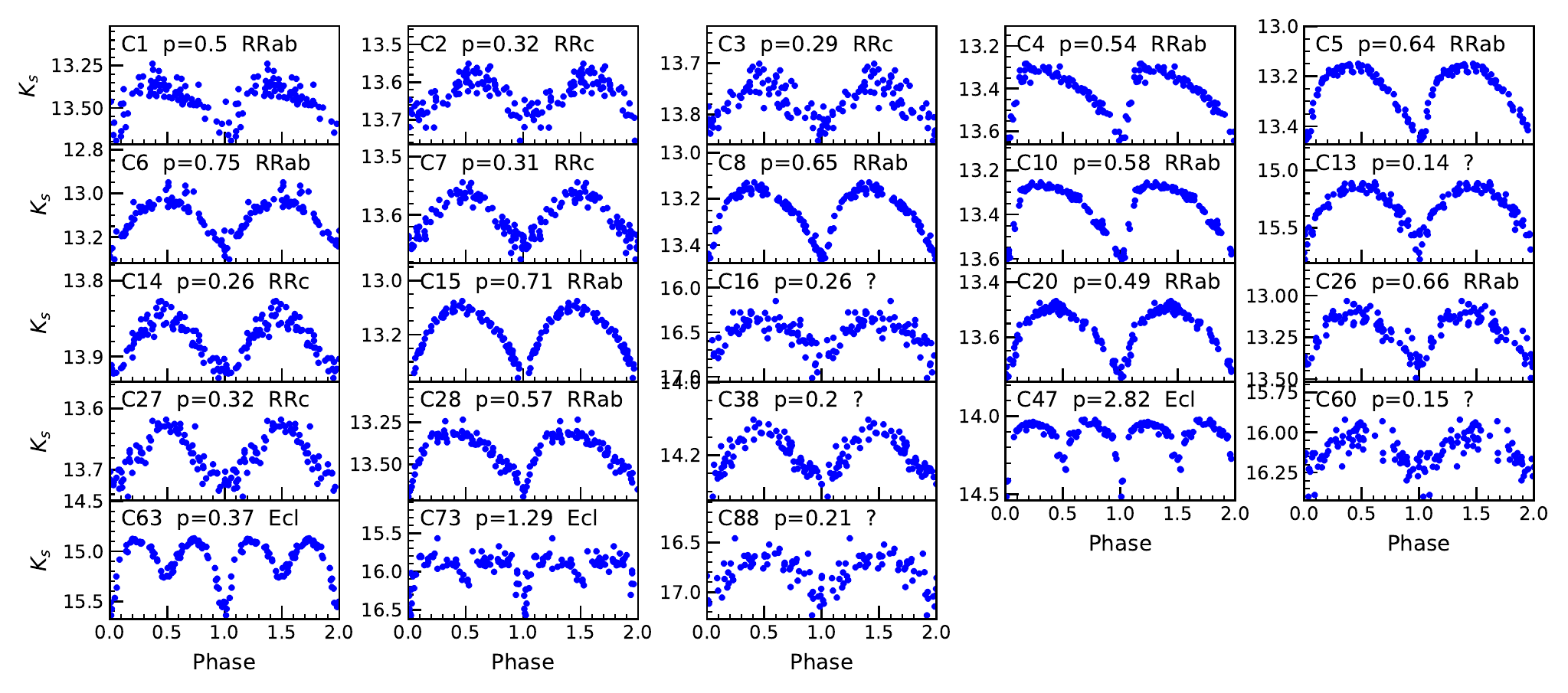}
  \caption{As in Fig.~\ref{fig_m22var}, but now for variable
    candidates in the M~28 field selected as cluster members, as shown
    in Sect.~\ref{sec_gc1pm}.}
  \label{fig_m28var}
\end{figure*}

\begin{figure*}
  \centering
  \includegraphics[scale=0.85]{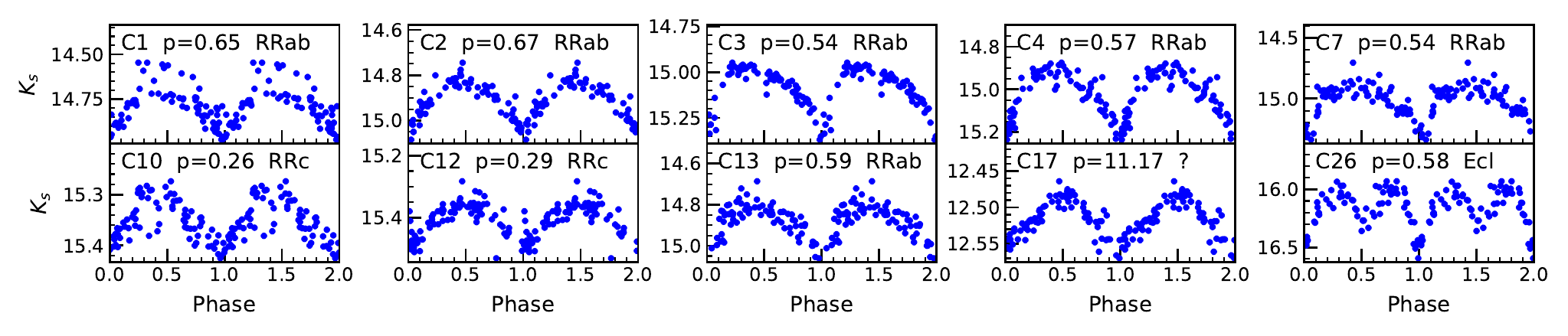}
  \caption{As in Fig.~\ref{fig_m22var}, but now for variable
    candidates in the NGC~6569 field selected as cluster members, as
    shown in Sect.~\ref{sec_gc1pm}.}
  \label{fig_ngc6569var}
\end{figure*}

\begin{figure*}
  \centering
  \includegraphics[scale=0.85]{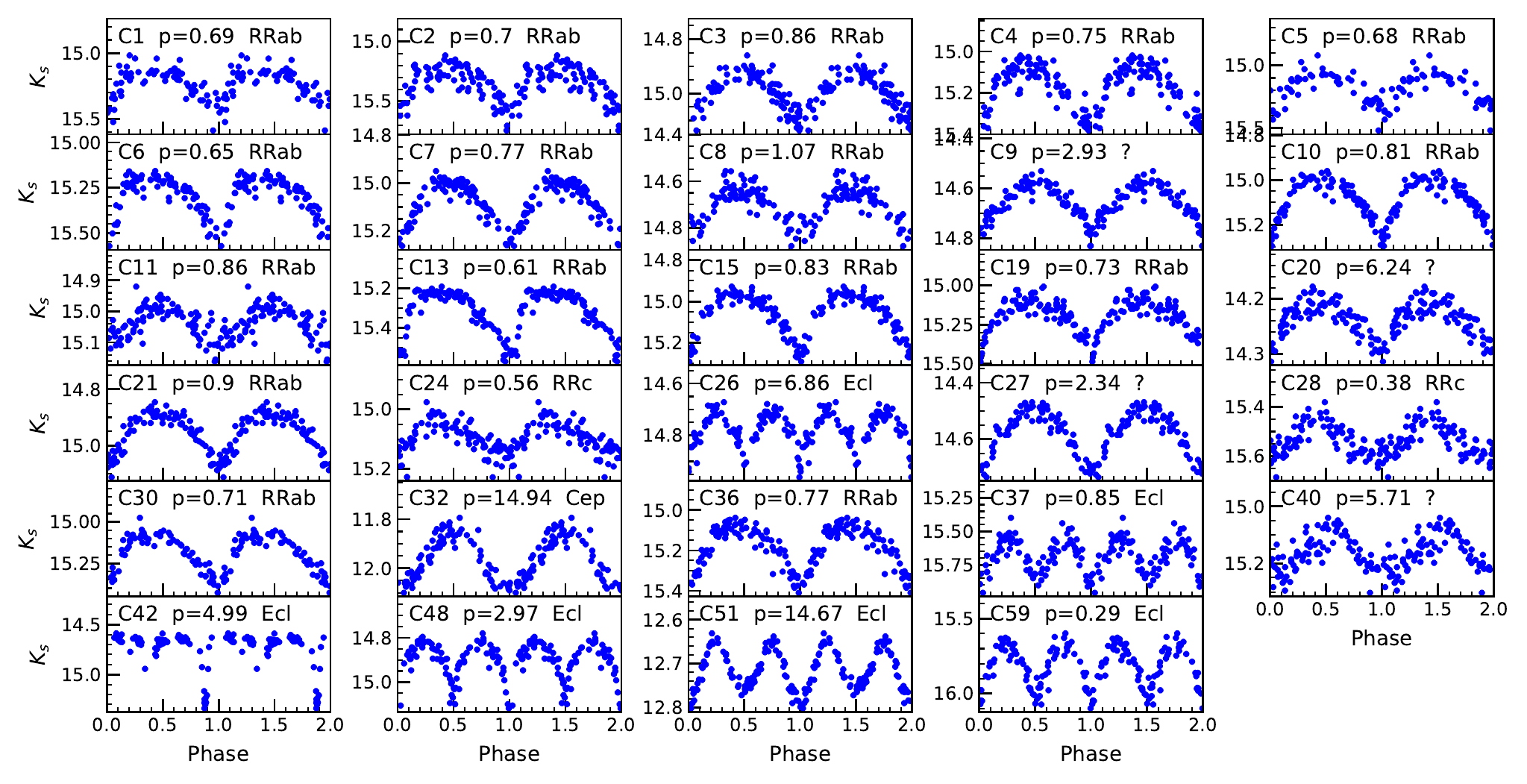}
  \caption{As in Fig.~\ref{fig_m22var}, but now for variable
    candidates in the NGC~6441 field selected as cluster members, as
    shown in Sect.~\ref{sec_gc1pm}.}
  \label{fig_ngc6441var}
\end{figure*}

In the end, based on their PMs and positions in the CMD, and limiting
ourselves to the variables stars we detected, we consider cluster
members 23 variable stars in M~28 (including 10 RRab and 5 RRc), 10 in
NGC~6569 (including 6 RRab and 2 RRc), and 29 in NGC~6441 (including
16 RRab, 2 RRc, and 1 Cepheid). They are identified as such in
Tables~\ref{tab_m28var}, \ref{tab_ngc6569var}, and
\ref{tab_ngc6441var}, respectively. Their corresponding light curves
are shown in Figs.~\ref{fig_m28var}, \ref{fig_ngc6569var}, and
\ref{fig_ngc6441var}.

\subsection{Distance and extinction}
\label{sec_gc1dis}
The GGCs in this section show a wide range of iron contents (see
Table~\ref{tab_gcinput}). Taking [${\alpha}$/Fe]=0.3 for all of the
GGCs \citep{vil17,joh18,ori08} translates into metallicities
$Z_{M28}$=0.0013, $Z_{NGC6569}$=0.0048, and $Z_{NGC6441}$=0.0096,
using Eq. (1) in \citet{nav17}. Applying the newly calibrated
\citetalias{alo15} PLZ relations from Sect.~\ref{sec_m22dis} (Eqs. [3]
to [7]) to the RR~Lyrae selected in Sect.~\ref{sec_gc1pm} as cluster
members (15 in M~28, 8 in NGC~6569, and 18 in NGC~6441), and assuming
a \citet{car89} canonical extinction law toward these as well, we
obtain the color excesses and distances for M~28, NGC~6569, and
NGC~6441 that we report in Table~\ref{tab_gcdis}.

The reported distances agree, within the error, with those reported by
\citet{bau19} using {\em Gaia} data, and also with those in the Harris
catalog (see Table~\ref{tab_gcdis}). NGC~6441 is the only case that
seems to be off ($\sim10\%$). This may imply that the uncommon
Oosterhoff~III GGCs follow different PLZ relations, a very plausible
posibility if we consider that the atypical extension of the HB and
the long periods in the RR~Lyrae in Oosterhoff~III GGCs may be
attributed to an overabundance of helium
\citep[e.g.,][]{cal07,cat09a,tai17}, which, as shown in \citet{cat04},
significantly impacts the PLZ relations that are followed by RR~Lyrae
stars.

\begin{figure*}
  \centering
  \begin{tabular}{ccc}
  \includegraphics[scale=0.48]{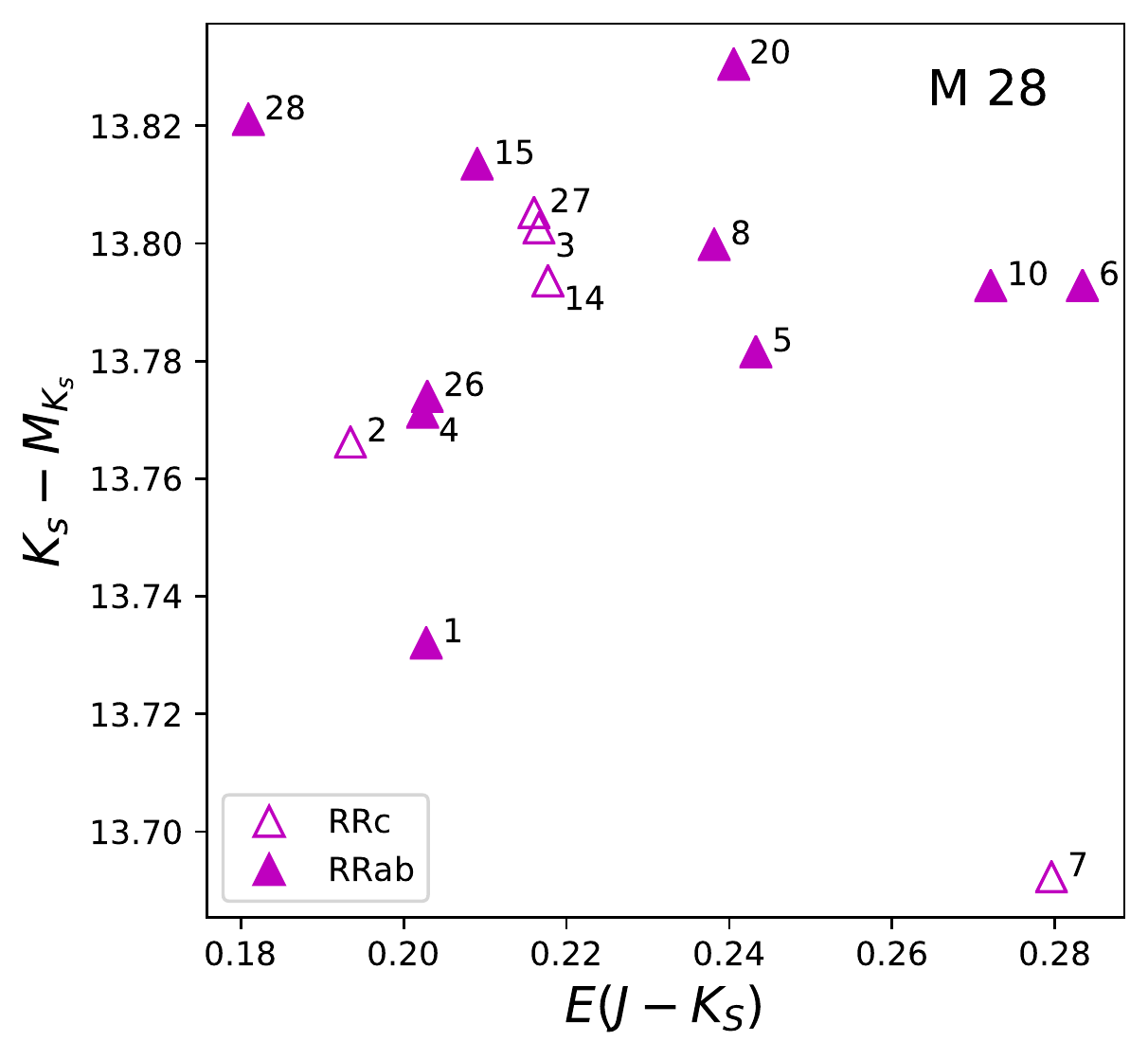}
  \includegraphics[scale=0.48]{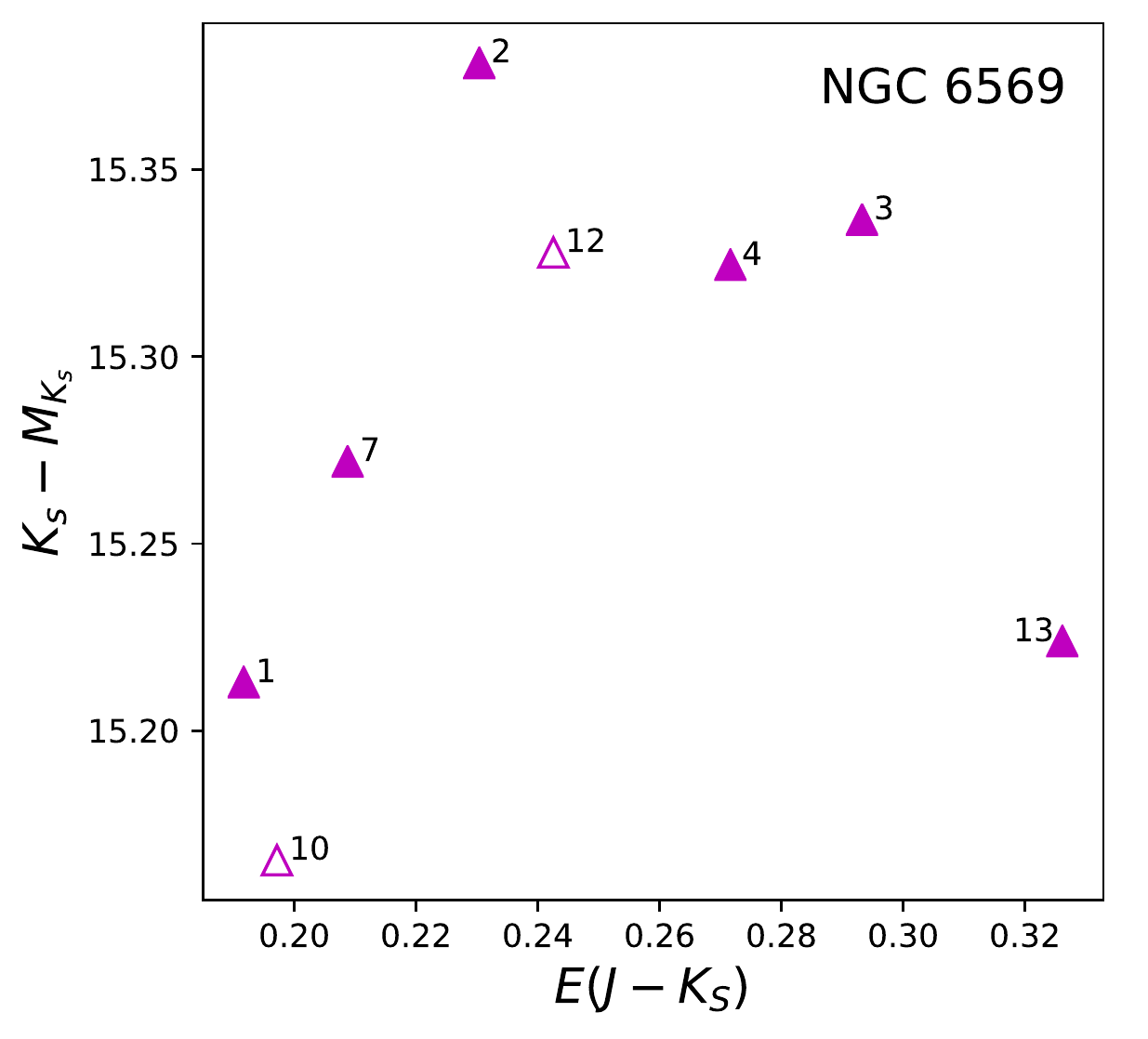}
  \includegraphics[scale=0.48]{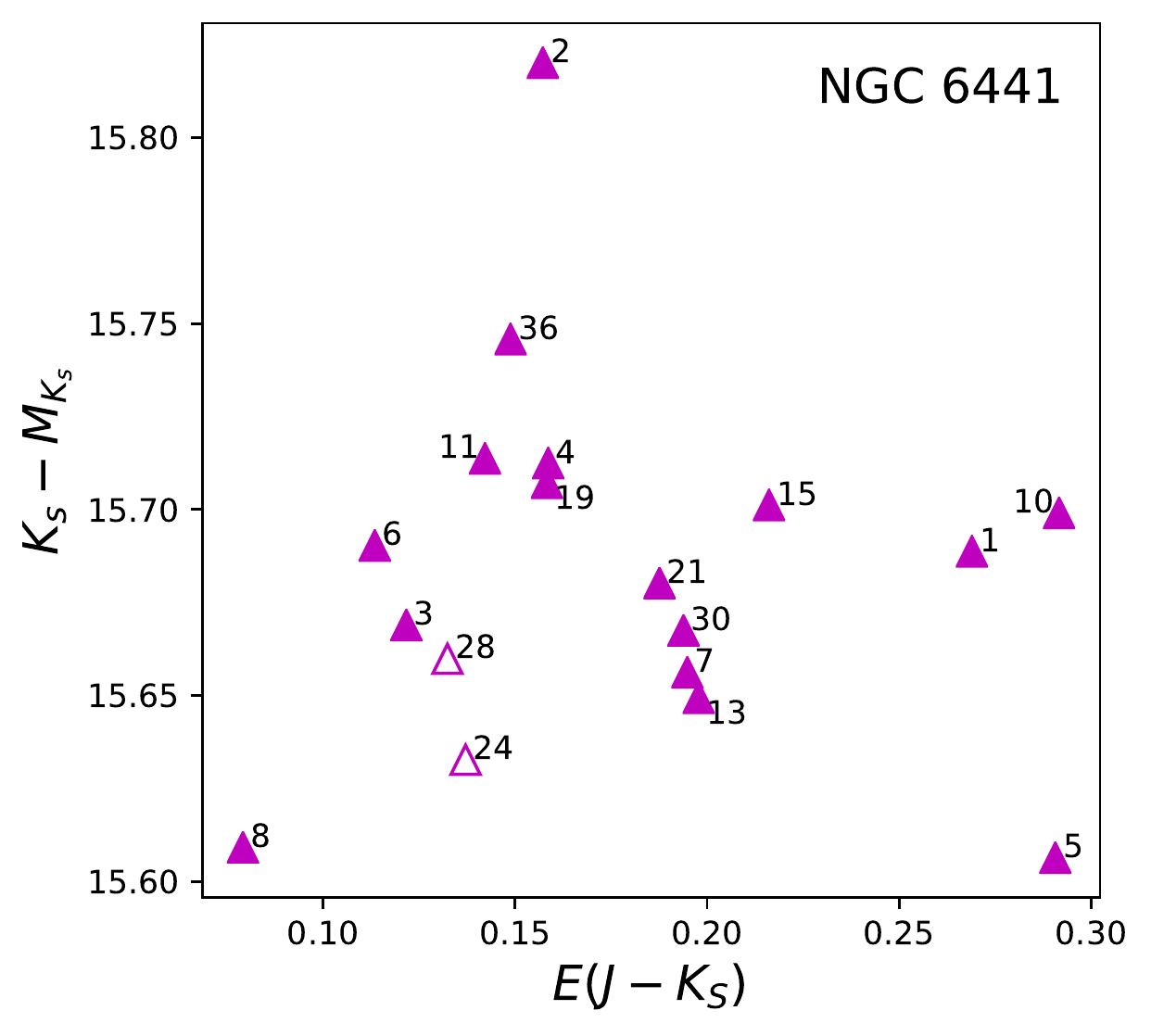}
  \end{tabular}
  \caption{As in Fig.~\ref{fig_m22dis}, but for NGC~6626 (M~28) in the
    left panel, for NGC~6569 in the middle panel, and for NGC~6441 in
    the right panel.}
  \label{fig_gc1dis}
\end{figure*}

The dispersion among the distance moduli to individual RR~Lyrae in
every GGC is small (see Fig.~\ref{fig_gc1dis}). It is unlikely to be
due to differential reddening because extinction is relatively low in
the near-infrared for these GGCs (see Table~\ref{tab_gcdis}), and the
dispersion among the stars does not seem to follow the reddening
vector (e.g., C28 for M~28 or C13 for NGC~6569 in
Fig.~\ref{fig_gc1dis}). This dispersion is quoted in
Table~\ref{tab_gcdis} as the first ${\sigma}$ in our distance
determination. Interestingly, this dispersion is similar for NGC~6441
as for the other GGCs.

We note that assuming a non-canonical extinction law such as those
reported for the innermost Galaxy
\citep[e.g.,][]{maj16,alo17,nog18,dek19} results in increasing the
distances to these GGCs. Variations in the distance due to applying
such extinction laws are included in the reported second $\sigma$ of
the distance estimation in Table~\ref{tab_gcdis}. We also note that
applying the different PLZ relations shown in Sect.~\ref{sec_m22dis}
only produces small variations among the reported mean ${\mu}_{Ks}$
for every cluster, increasing their values from a few hundredths of a
magnitude, in the case of the newly calibrated \citet{nav17b} PLZ
relations (Eqs. [1] and [2] in Sect.~\ref{sec_m22dis}), up to
$\sim0.1$ mag, for the \citet{mur15} PLZ relation. Variations in the
distance due to applying these different PLZ relations are included in
the reported third $\sigma$ of the distance estimation in
Table~\ref{tab_gcdis}.

There are no Cepheids in M~28 and NGC~6569 that we detected as cluster
members, but there is one, C34, in NGC~6441. When we examined its
colors, we found it was saturated in $H$ but not in $J$. Using the PL
relations by \citet{dek19}, we found a color excess $E(J-K_s)=0.27$
for this Cepheid, and a distance of $R_{\odot}=13.3$ kpc, in agreement
with those reported by the RR~Lyrae for this GGC (see
Table~\ref{tab_gcdis}), but significantly higher than that reported by
{\em Gaia} and in the Harris catalog. Therefore, this result hints
that a different calibration for the PL relations in Cepheids from
Oosterhoff~III GGCs may also be required.

\section{2MASS-GC~02 and Terzan~10}
\label{sec_gc2}
2MASS-GC~02 and Terzan~10 are different from the previously analyzed
GGCs because they are two poorly populated GGCs located at very low
latitudes (see Table~\ref{tab_gcinput}). Their Galactic locations make
them two of the most reddened GGCs in the Milky Way. We studied the
variable populations in these GGCs and in their immediate surroundings
in \citetalias{alo15}. Now we revisit these two GGCs as test cases of
our redesigned method to detect variables.

\subsection{Variables in the cluster area}
\label{sec_gc2var}
2MASS-GC~02 is one of the very few Galactic GGCs lying in the
Oosterhoff gap, as we showed in \citetalias{alo15}. We detected 45
variable candidates in its field, following the procedures detailed in
Sect.~\ref{sec_var}. We present their main observational parameters in
Table~\ref{tab_2msgc2var}. Owing to its position, very close to the
Galactic plane, this GGC is outside the footprint of the OGLE
experiment. There are 6 variable candidates listed in the Clement
catalog, but we discarded these as not being real variables in
\citetalias{alo15}. We do not register any signal of variability for
these stars in our current analysis either. In \citetalias{alo15} we
were able to detect 32 variables inside the tidal radius of this
GGC. We independently recovered all the variable stars detected in
\citetalias{alo15}, except those classified as long period variables
(LPVs): NV16, NV19, and NV27; and the eclipsing binary NV32. This is
expected as we explained in Sect.~\ref{sec_var}: Our method is now
restricted to looking for stars with periods shorter than 100 days,
and therefore we do not expect to find any LPVs; as for the eclipsing
binaries, we expect to miss some of these because of the preliminary
cuts we apply to our selection. We underline that we recovered all 16
pulsators (13 RRab's and 3 Cepheids) previously found in
\citetalias{alo15} inside the tidal radius of this GGC. Additionally,
we note that the coordinates for the cluster center we used in this
work (see Table~\ref{tab_gcinput}) are slightly different from those
assumed in \citetalias{alo15}, which puts the Cepheid candidate C45
(NV33) inside the tidal radius, and among our detected variables. On
the other hand, however, the improved light curve of C40 (NV28) casts
some doubts on our previous classification as a Cepheid, and we prefer
not to assign this source a variable type. Finally, we highlight that
there are 16 newly detected candidates. We classified 2 as RRab, 4 as
eclipsing binaries, and we were not able to assign a variable type to
the other 10 new candidates.

\begin{table*}
  \caption{Properties of the variable candidates in 2MASS-GC~02.}
  \label{tab_2msgc2var}
  \centering
  \resizebox{\textwidth}{!}{
  \begin{tabular}{cccccccccccccccc}
    \hline\hline
    ID & $ID_{PaperI}$ & $\alpha$ (J2000) & ${\delta}$ (J2000) & Distance\tablefootmark{a} & Period & $A_{K_s}$ & $\langle K_s \rangle$ & $Z-K_s$ & $Y-K_s$ & $J-K_s$ & $H-K_s$ & $\mu_{\alpha^\ast}$ & $\mu_{\delta}$ & Member\tablefootmark{b} & Type \\
    & & (h:m:s) & (d:m:s) & (arcmin) & (days) & (mag) & (mag)& (mag)& (mag)& (mag) & (mag) & (mas/yr) & (mas/yr) & & \\ 
    \hline
C1 & NV1 & 18:09:35.98 & -20:47:11.7 & 0.52 & 0.70004 & 0.298 & 14.585 & -- & 4.411 & 2.783 & 0.95 & 4.89 & -4.81 & Yes & RRab\\
C2 & NV2 & 18:09:34.22 & -20:46:56.2 & 0.68 & 0.651668 & 0.218 & 14.779 & -- & -- & 2.942 & 1.008 & 3.785 & -4.757 & Yes & RRab\\
C3 & NV4 & 18:09:38.58 & -20:46:05.0 & 0.75 & 0.623721 & 0.243 & 15.356 & -- & -- & 4.176 & 1.494 & 4.421 & -2.921 & Yes & RRab\\
C4 & NV3 & 18:09:33.77 & -20:46:29.5 & 0.79 & 0.570441 & 0.279 & 15.069 & -- & -- & 3.46 & 1.155 & 5.01 & -3.961 & Yes & RRab\\
C5 & NV5 & 18:09:38.85 & -20:47:26.5 & 0.83 & 0.603305 & 0.33 & 14.973 & -- & -- & 3.213 & 1.099 & 3.846 & -3.893 & Yes & RRab\\
C6 & NV6 & 18:09:33.00 & -20:47:07.8 & 1.02 & 0.551335 & 0.404 & 14.965 & -- & -- & 2.937 & 1.004 & 3.038 & -6.488 & Yes & RRab\\
C7 & -- & 18:09:38.86 & -20:45:46.7 & 1.05 & 0.60917 & 0.23 & 15.557 & -- & -- & -- & 1.758 & 2.888 & -2.794 & Yes & RRab\\
C8 & -- & 18:09:33.46 & -20:47:32.9 & 1.16 & 1.068307 & 0.081 & 14.747 & 3.327 & 2.277 & 1.412 & 0.503 & -2.931 & -0.364 & No & ?\\
C9 & -- & 18:09:40.77 & -20:47:43.5 & 1.33 & 3.2458 & 0.271 & 14.507 & 3.734 & 2.591 & 1.594 & 0.537 & 0.486 & -1.929 & No & Ecl\\
C10 & -- & 18:09:31.65 & -20:47:24.7 & 1.42 & 0.596172 & 0.34 & 14.687 & -- & -- & 2.924 & 1.018 & 3.094 & -5.215 & Yes & RRab\\
    \hline
  \end{tabular}}
\tablefoot{This table is available in its entirety in electronic form
  at the CDS. A portion is shown here for guidance regarding its form and content.\\
\tablefoottext{a}{Projected distance to the cluster center}
\tablefoottext{b}{Cluster membership according to criteria explained in Sect.~\ref{sec_gc2dis}}
}
\end{table*}

Terzan~10 shows in \citetalias{alo15} mean periods for its RRab star
corresponding to an Oosterhoff~II GGC. But its reported metallicity at
the time, from photometric measurements, $[{\rm Fe/H}]=-1.0$ according
to the Harris catalog, was too high for this group, putting it closer
to the Oosterhoff~III GGCs. As shown in Table~\ref{tab_gcinput}, in
this paper we use a lower iron content, $[{\rm Fe/H}]=-1.59\pm0.02$,
which is measured from Ca triplet spectra in a sample of 16 member
stars observed with the FORS2 instrument at the Very Large
Telescope (VLT; Geisler et al., in prep.). This metallicity is more
consistent with Terzan~10 being a member of the Oosterhoff~II GGCs. We
detected 65 variable candidates in its field. We present their main
observational parameters in Table~\ref{tab_terzan10var}. The Clement
catalog for this GGC shows the variable stars we detected in
\citetalias{alo15}, plus a few more candidates from OGLE. There are 48
variable stars listed in \citetalias{alo15} inside the tidal radius of
Terzan~10. We recovered all but 7 of these variables: 1 LPV (NV48),
since our method is now restricted to looking for stars with periods
shorter than 100 days (see Sect.~\ref{sec_var}); and the other six -- 3
variable sources of unknown type (NV8, NV37, and NV38), 2 Cepheid
candidates (NV19 and NV39), and 1 eclipsing binary candidate (NV45)--,
whose light curves quality was too low to be picked out by our
improved method. We note that, as for 2MASS-GC~02, the coordinates for
the cluster center we used in this work (see Table~\ref{tab_gcinput})
are slightly different from those assumed in \citetalias{alo15} for
this GGC, which puts NV50 and NV51 from \citetalias{alo15} inside its
tidal radius, and among our detected variables. We also reconsidered
our previous classifications for some of these stars based on their
improved light curves: we classified C56 (NV44) as an RRab, while we
were not able to assign a variable type to all the 6 stars (C15, C16,
C18, C19, C47, and C54) previously classified as Cepheid candidates
(NV16, NV18, NV14, NV17, NV31, and NV47). Finally, there are 22
variable candidates not reported in \citetalias{alo15} that we
detected in this work. From those, OGLE previously detected 12
variable stars and the other 10 are newly detected candidates. We
classified 2 of them as eclipsing binaries, and we were not able to
assign a variable type to the other 8 new candidates.

\begin{table*}
  \caption{Properties of the variable candidates in Terzan~10.}
  \label{tab_terzan10var}
  \centering
  \resizebox{\textwidth}{!}{
  \begin{tabular}{ccccccccccccccccc}
    \hline\hline
    ID & $ID_{Clement}$ & $ID_{OGLE}$ & $\alpha$ (J2000) & ${\delta}$ (J2000) & Distance\tablefootmark{a} & Period & $A_{K_s}$ & $\langle K_s \rangle$ & $Z-K_s$ & $Y-K_s$ & $J-K_s$ & $H-K_s$ & $\mu_{\alpha^\ast}$ & $\mu_{\delta}$ & Member\tablefootmark{b} & Type \\
    & & (OGLE-BLG-) & (h:m:s) & (d:m:s) & (arcmin) & (days) & (mag) & (mag)& (mag)& (mag)& (mag) & (mag) & (mas/yr) & (mas/yr) & & \\ 
    \hline
C1 & V1 & ECL-270713 & 18:02:58.87 & -26:03:35.3 & 0.46 & 3.87962 & 0.39 & 14.14 & 2.713 & 1.833 & 1.069 & 0.31 & -0.118 & -9.19 & No & Ecl\\
C2 & V2 & RRLYR-33526 & 18:02:59.48 & -26:04:22.6 & 0.5 & 0.730538 & 0.291 & 14.655 & 2.734 & 1.831 & 1.105 & 0.331 & -6.908 & -4.292 & Yes & RRab\\
C3 & V3 & RRLYR-33512 & 18:02:54.04 & -26:03:46.9 & 0.92 & 0.70172 & 0.298 & 14.842 & 3.17 & 2.005 & 1.154 & 0.383 & -7.334 & -1.237 & Yes & RRab\\
C4 & V4 & ECL-270954 & 18:03:00.19 & -26:05:05.8 & 1.2 & 1.368958 & 0.395 & 14.651 & 2.082 & 1.421 & 0.798 & 0.249 & -2.839 & -6.769 & No & Ecl\\
C5 & V5 & -- & 18:02:57.14 & -26:02:44.0 & 1.28 & 0.68854 & 0.314 & 14.974 & 3.571 & 2.522 & 1.338 & 0.405 & -6.43 & -1.814 & Yes & RRab\\
C6 & V6 & RRLYR-33519 & 18:02:56.90 & -26:05:19.7 & 1.35 & 0.582332 & 0.334 & 14.957 & 3.199 & 2.01 & 1.14 & 0.361 & -6.184 & -2.43 & Yes & RRab\\
C7 & V7 & RRLYR-33509 & 18:02:53.24 & -26:05:12.7 & 1.62 & 0.715331 & 0.339 & 14.726 & 2.928 & 1.915 & 1.147 & 0.409 & -8.985 & -4.512 & Yes & RRab\\
C8 & V10 & ECL-271769 & 18:03:04.65 & -26:05:15.4 & 1.95 & 0.422508 & 0.45 & 16.306 & 2.033 & 1.251 & 0.768 & 0.221 & -4.449 & -0.437 & Yes & Ecl\\
C9 & V9 & -- & 18:02:51.22 & -26:05:14.9 & 1.97 & 0.193836 & 0.236 & 16.207 & 2.325 & 1.605 & 0.921 & 0.239 & -5.788 & -4.402 & Yes & ?\\
C10 & V12 & RRLYR-33538 & 18:03:04.08 & -26:05:33.7 & 2.07 & 0.568658 & 0.291 & 14.533 & 2.408 & 1.7 & 0.939 & 0.265 & -1.84 & -7.808 & No & RRab\\
    \hline
  \end{tabular}}
\tablefoot{This table is available in its entirety in electronic form
  at the CDS. A portion is shown here for guidance regarding its form and content.\\
\tablefoottext{a}{Projected distance to the cluster center}
\tablefoottext{b}{Cluster membership according to criteria explained in Sect.~\ref{sec_gc2dis}}
}
\end{table*}

Therefore, from the analysis of these 2 difficult GGCs, we note that
our redesigned method to detect variable candidates (see
Sect.~\ref{sec_var}) recovered all variable stars detected in
\citetalias{alo15}, except for those with periods outside of our range
of interest and those with low quality light curves. On the other
hand, it increased the number of detected candidates with good quality
light curves by more than 50 percent. However, we also note that the
number of epochs available for our variability analysis was
significantly increased with respect to \citetalias{alo15}, by a
factor of $\sim2$ in 2MASS-GC~02 and a factor of $\sim3$ in Terzan~10.

\subsection{Proper motions, color-magnitude diagrams, and cluster memberships}
\label{sec_gc2pm}
\begin{figure*}
  \centering
  \begin{tabular}{cc}
  \includegraphics[scale=0.37]{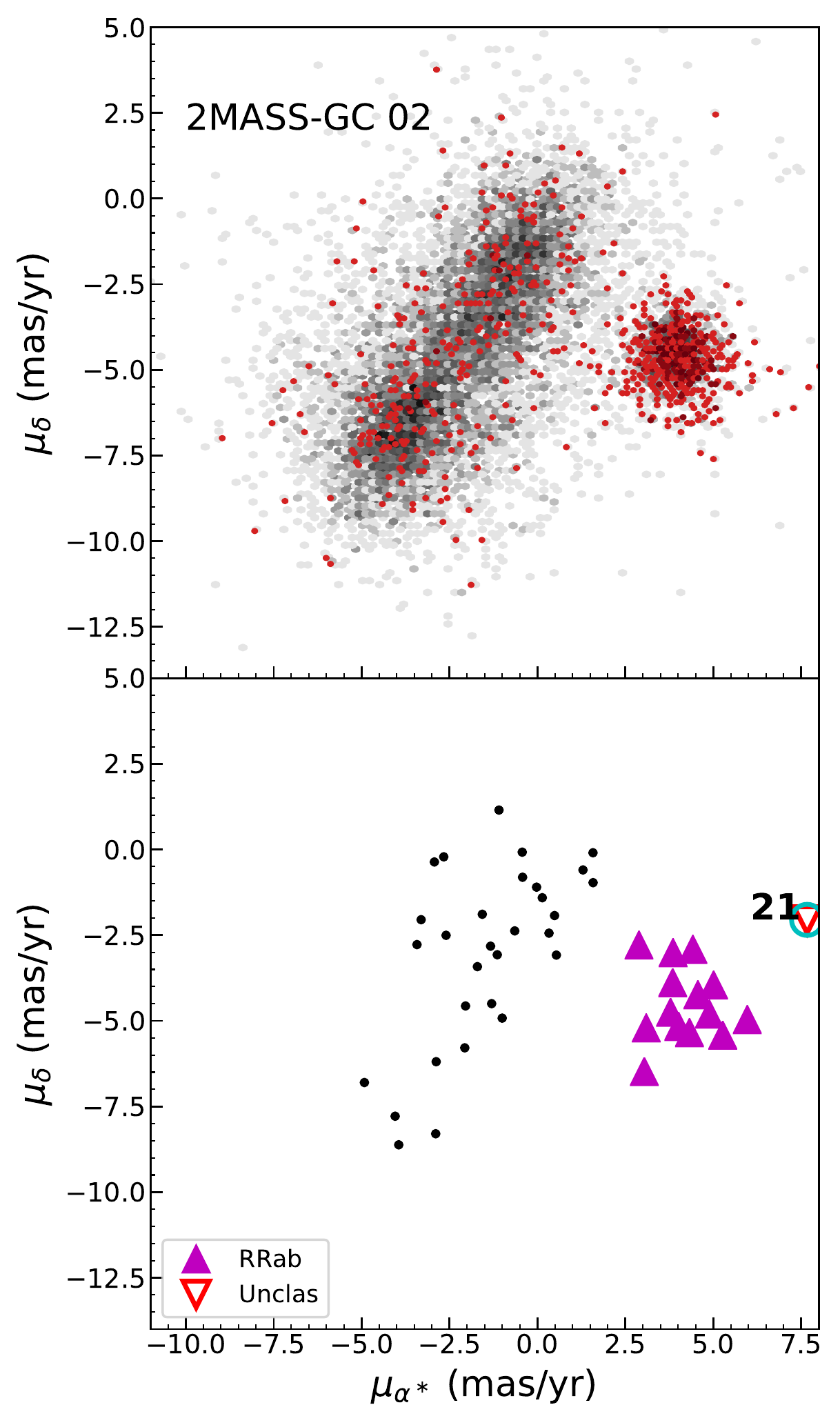}
  \includegraphics[scale=0.37]{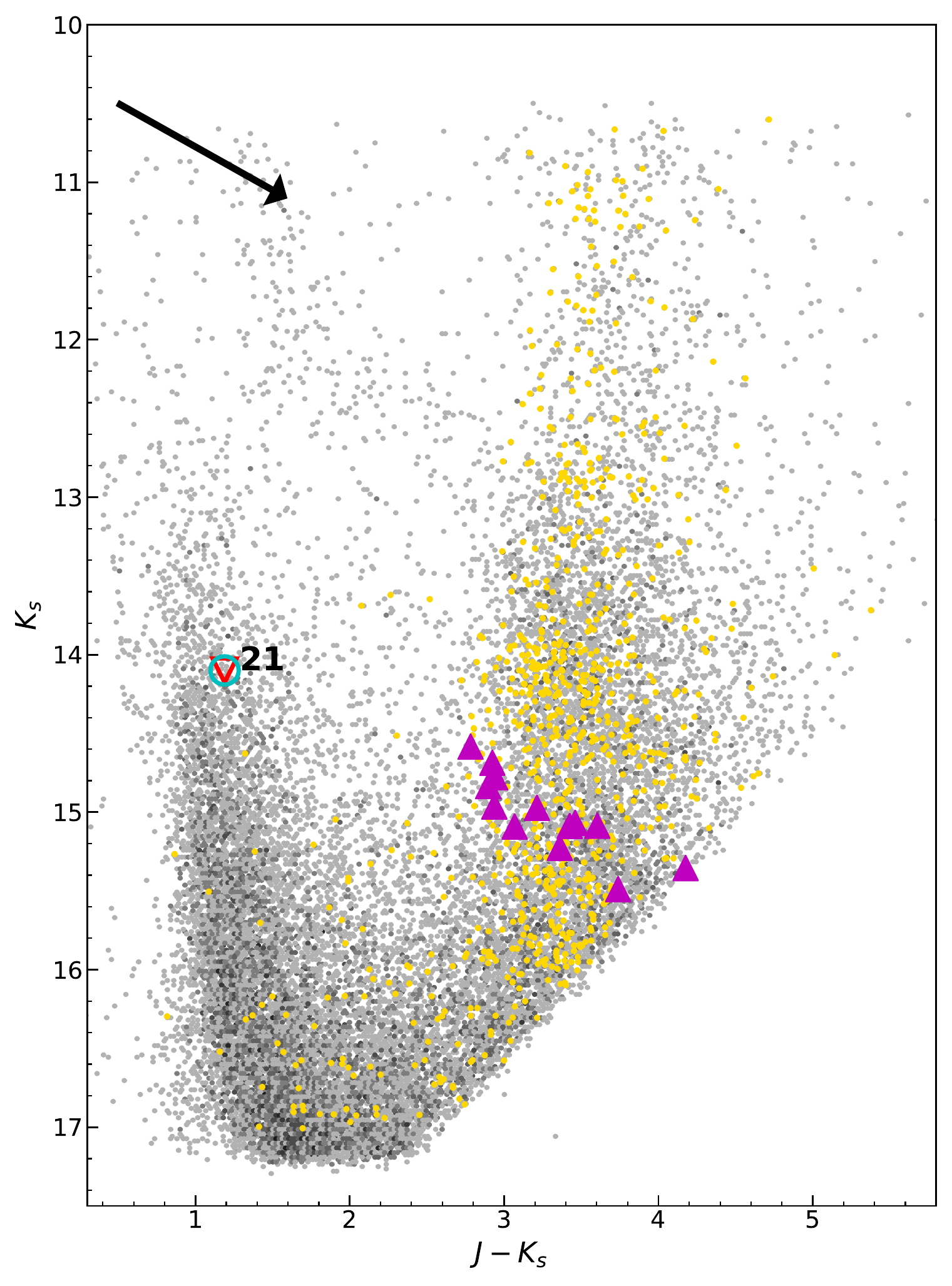}\\
  \includegraphics[scale=0.37]{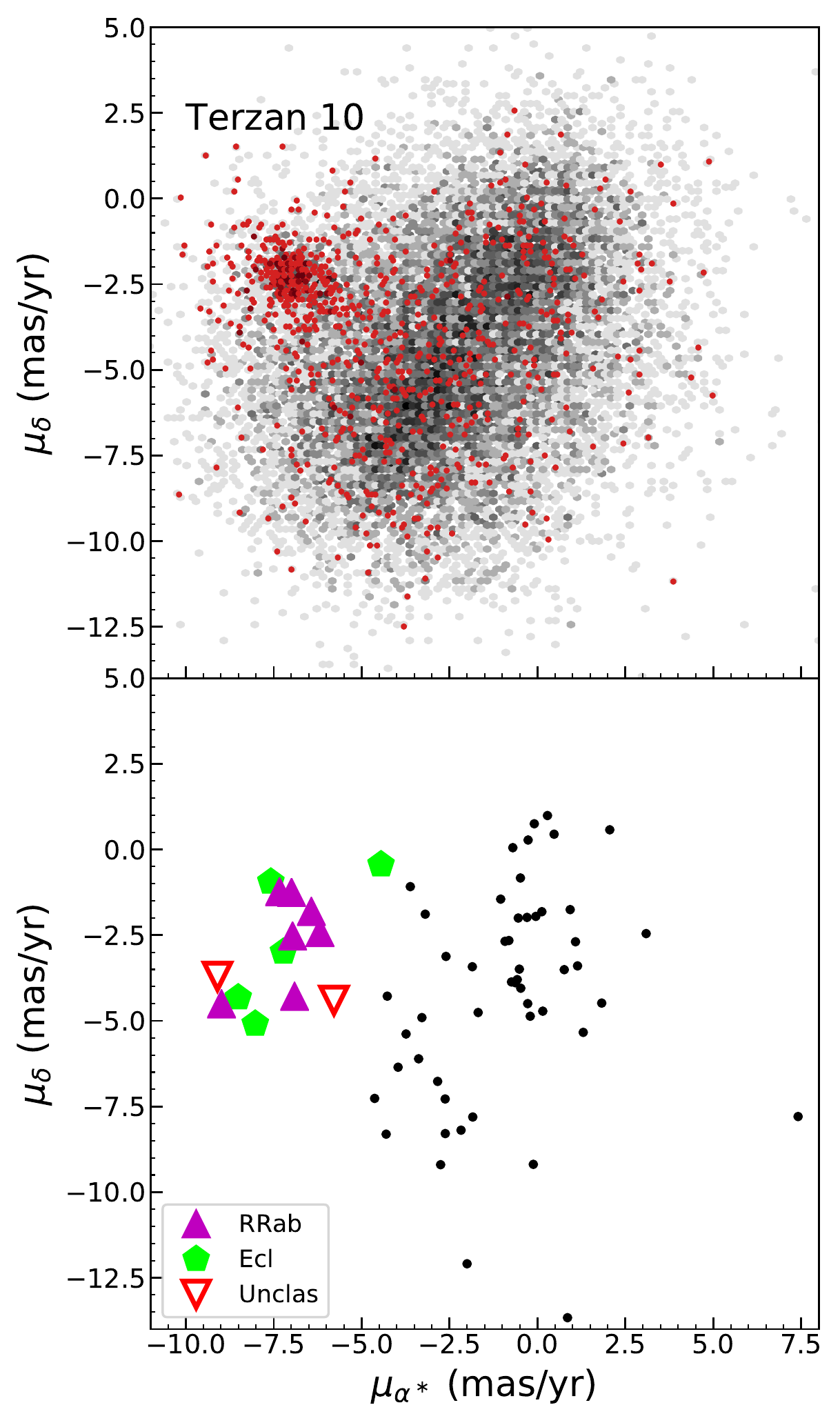}
  \includegraphics[scale=0.37]{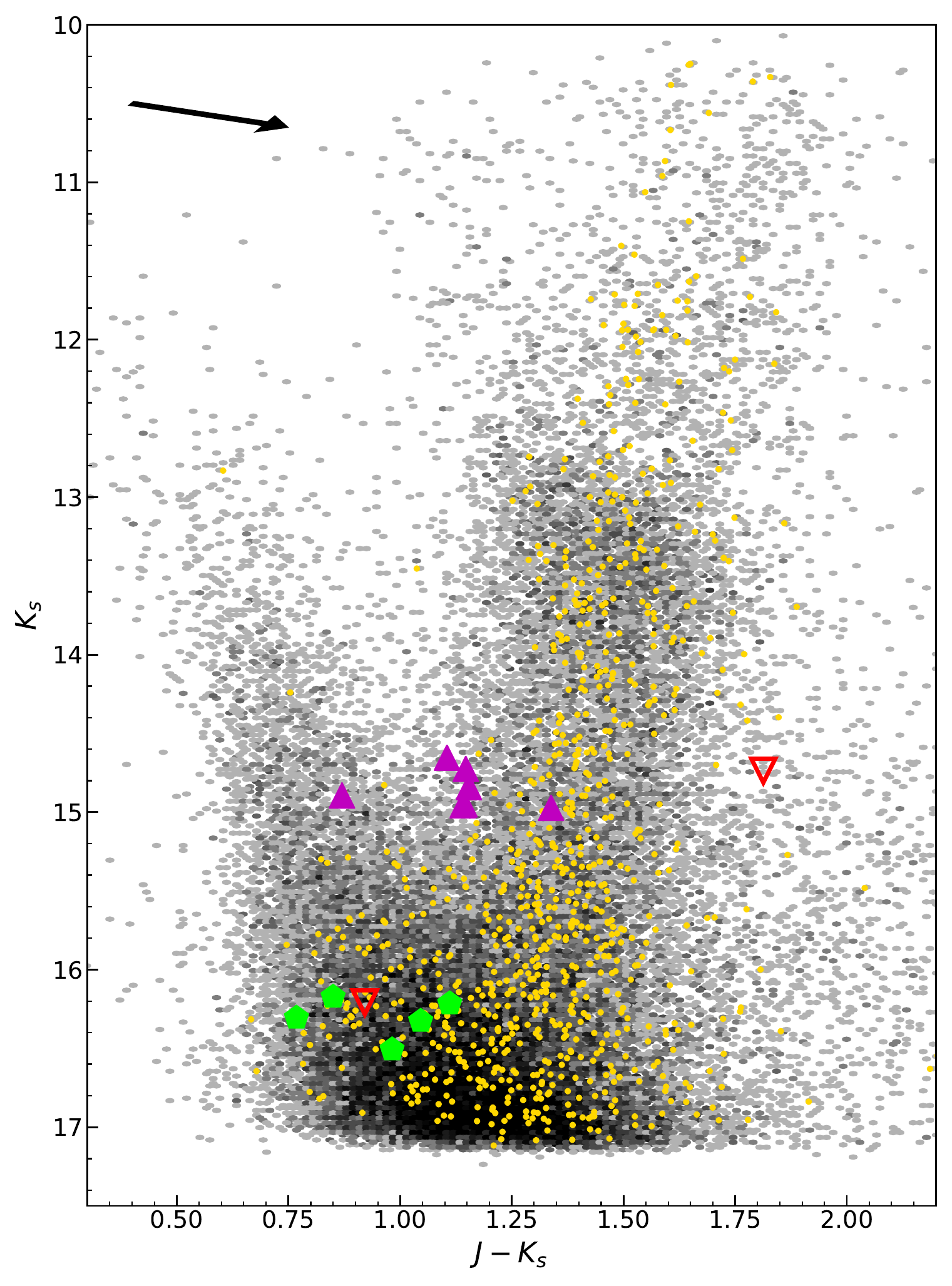}\\
  \end{tabular}
  \caption{As in Fig.~\ref{fig_m22}, but now for 2MASS-GC~02 in the
    top panels and for Terzan~10 in the lower panels. For 2MASS-GC~02,
    a cyan empty circle encapsulates the variable candidate whose
    membership to the cluster according to our kNN classifier was
    reversed (see text).}
  \label{fig_gc2}
\end{figure*}

Following the same method described in Sect.~\ref{sec_m22pm} and
\ref{sec_gc1pm}, we used the PMs provided by VIRAC2 for the detected
variable stars to assign membership to 2MASS-GC~02 and Terzan~10. In
the left upper panels of Fig.~\ref{fig_gc2}, we can observe that the
PM distributions of stars from both GGCs are more clearly separated
from their field counterparts than in the GGCs from the previous
section (see Fig.~\ref{fig_gc1}). Our kNN classifier allowed us to
further select the innermost stars that belong to the GGCs. We are
able to obtain in this way the PMs for 2MASS-GC~02 and Terzan~10, which
we show in Table~\ref{tab_gcpm}. While the mean PM for Terzan~10
agrees with that measured by \citet{bau19}, this is not the case for
2MASS-GC~02. We attribute this to the very high reddening in front of
2MASS-GC~02, which may have produced a wrong identification of its
members using {\em Gaia} data by \citet{bau19}. We argue that our VVV
near-infrared data for this GGC provides more accurate results in this
case because a PM pattern for the stars in this GGC, different from
the stars in the field, and not observed in the {\em Gaia} data,
appears clearly in the left panels of Fig.~\ref{fig_gc2} for
2MASS-GC~02.

\begin{figure*}
  \centering
  \includegraphics[scale=0.85]{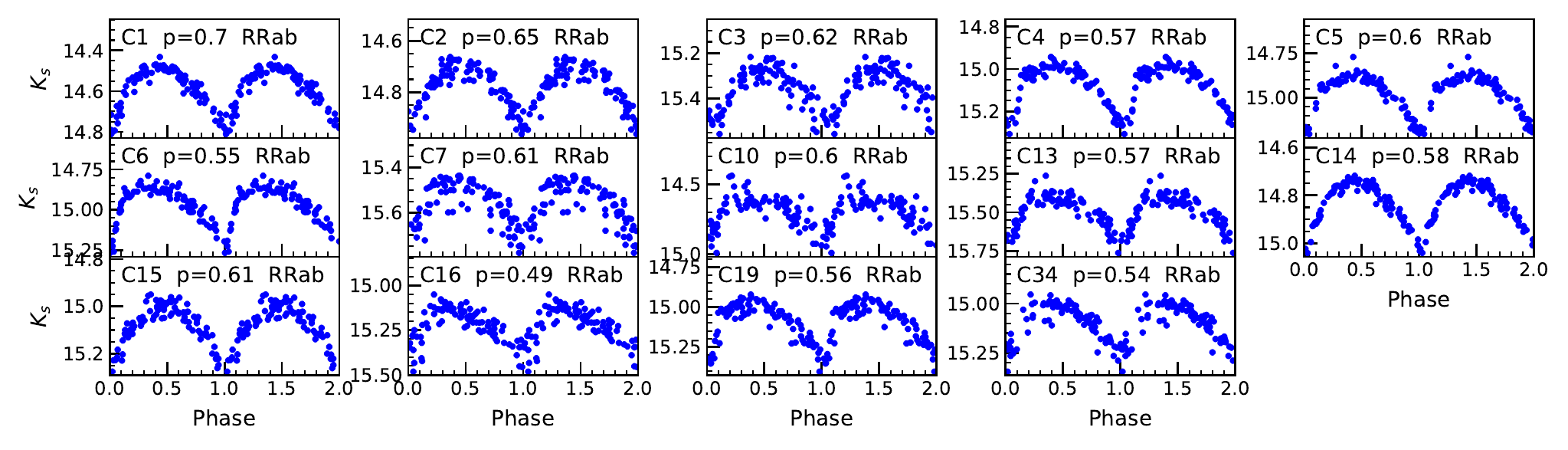}
  \caption{As in Fig.~\ref{fig_m22var}, but now for variable
    candidates in the 2MASS-GC~02 field selected as cluster members,
    as shown in Sect.~\ref{sec_gc2pm}.}
  \label{fig_2msgc2var}
\end{figure*}

\begin{figure*}
  \centering
  \includegraphics[scale=0.85]{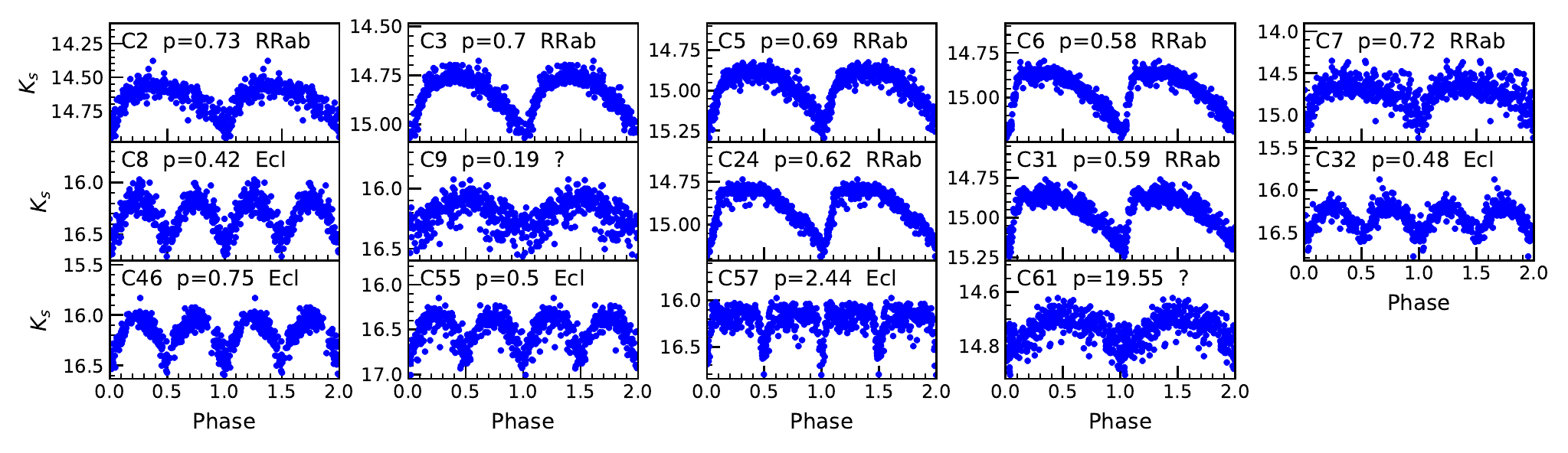}
  \caption{As in Fig.~\ref{fig_m22var}, but now for variable
    candidates in the Terzan~10 field selected as cluster members, as
    shown in Sect.~\ref{sec_gc2pm}.}
  \label{fig_terzan10var}
\end{figure*}

Our kNN classifier also allows us to identify cluster member variable
stars using their PMs (see left lower panels of
Fig.~\ref{fig_gc2}). There are 15 variable stars identified as members
for 2MASS-GC~02, all of which are RRab stars, except for 1
unclassified variable (C21). However, as the position of C21 in the
CMD is too blue to belong to this GGC and better agrees with a star in
the foreground disk (see top right panel in Fig.~\ref{fig_gc2}), we
decided to consider it a non-member in its classification in
Table~\ref{tab_2msgc2var}. On the other hand, all the RRab classified
as cluster members consistently fall along the reddening vector in the
CMD, indicating the presence of strong differential reddening in the
line of sight of this GGC. All the RRab selected as members in
\citetalias{alo15} were also selected in this work using the kNN
classifier. We present their $K_s$ light curves in
Fig.~\ref{fig_2msgc2var}. When also considering the additional 2 newly
detected RR~Lyrae in this GGC, the Oosterhoff intermediate
classification we assigned it in \citetalias{alo15} is reinforced. For
Terzan~10, our kNN classifier identifies 14 variable stars as cluster
members based on their PMs. We present their $K_s$ light curves in
Fig.~\ref{fig_terzan10var}. As for 2MASS-GC~02, all the RRab stars
classified in \citetalias{alo15} as cluster members are also
classified as members now.

\subsection{Distance and extinction}
\label{sec_gc2dis}
\begin{figure*}
  \centering
  \begin{tabular}{ccc}
  \includegraphics[scale=0.48]{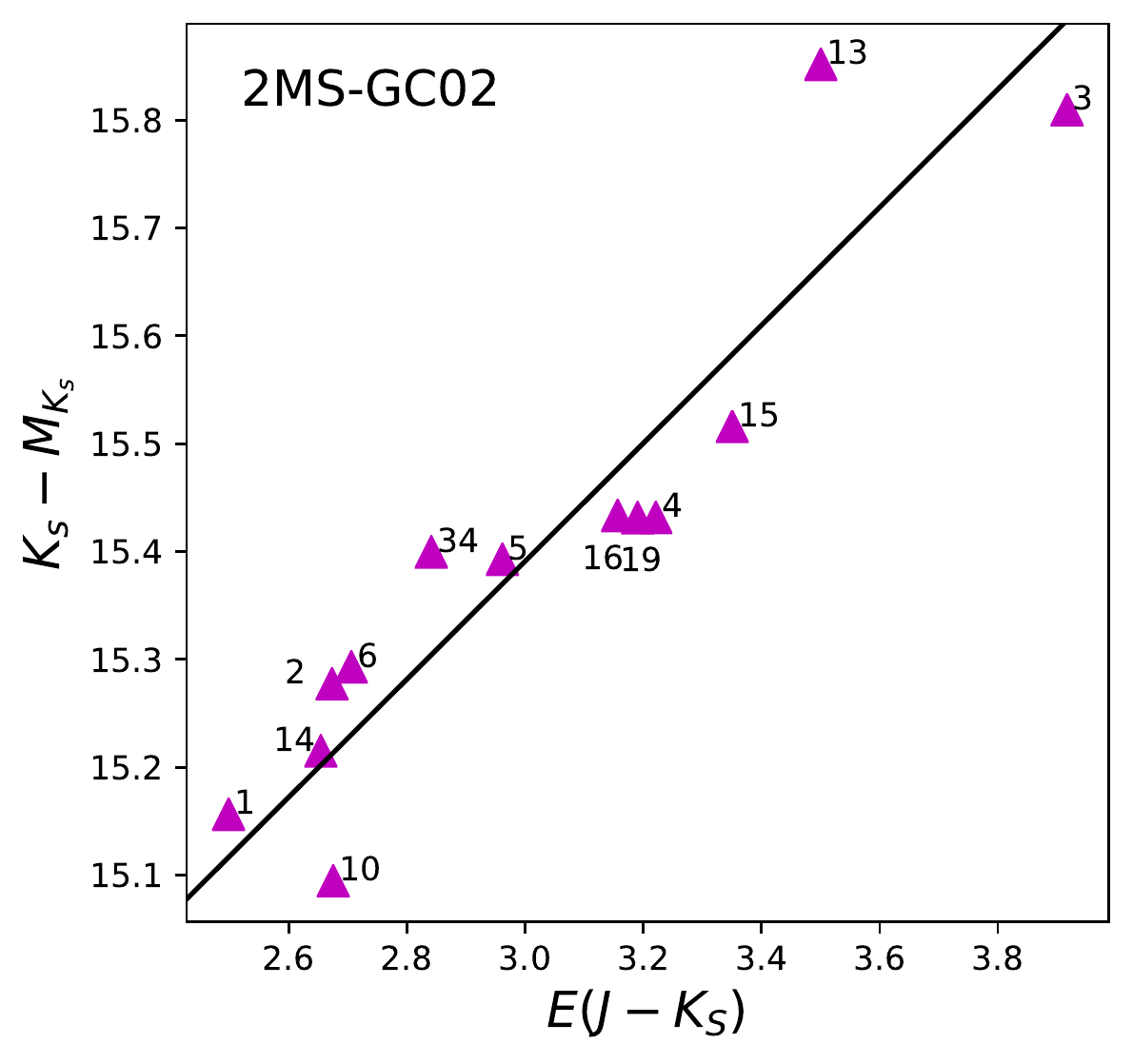}
  \includegraphics[scale=0.48]{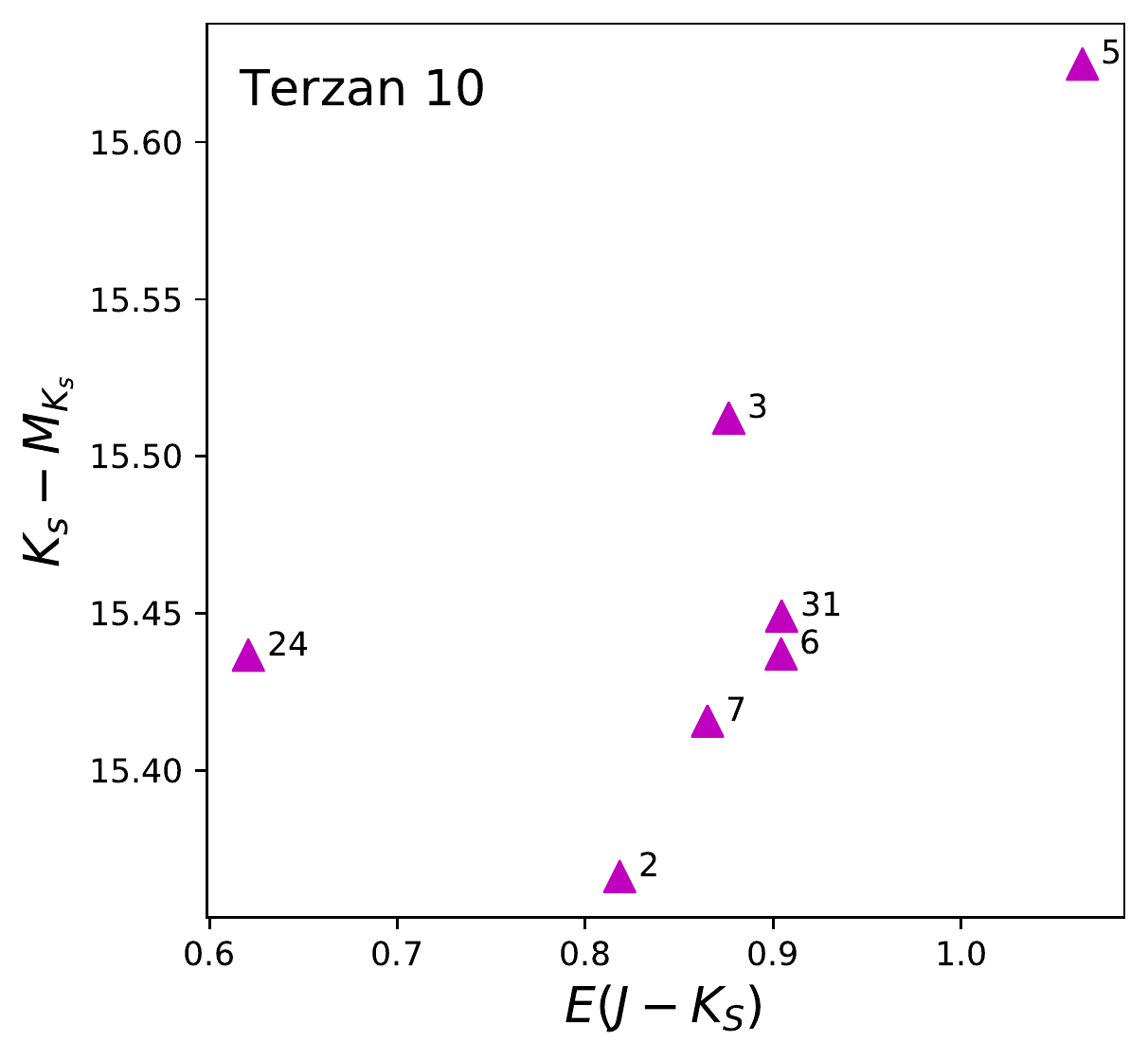}
  \end{tabular}
  \caption{As in Fig.~\ref{fig_m22dis}, but now for 2MASS-GC~02 in the
    left panel and for Terzan~10 in the right panel. Also shown as a
    straight line for 2MASS-GC~02 is the linear fit used to define its
    distance modulus corrected by extinction, as the zero-point of the
    fit, and its selective-to-total extinction ratio, as the slope of
    the fit.}
  \label{fig_gc2dis}
\end{figure*}

For our determination of distances and extinctions to 2MASS-GC~02, we
used 14 RRLyrae: the 12 cluster members detected in \citetalias{alo15}
(C1 to C6, C13 to C16, C19, and C34), plus 2 additional new detections
from our current analysis (C7 and C10), which our kNN classifier also
considers members of 2MASS-GC~02. The remaining RRab, C38, and the 3
detected Cepheids (C17, C43 and C45) were classified as field stars in
\citetalias{alo15} and in this work by our kNN classifier. As
mentioned in Sect.~\ref{sec_gc2pm}, the positions of the RRab stars of
2MASS-GC~02 in the CMD (see top right panel in Fig.~\ref{fig_gc2})
provided us with a clear hint of the significant differential
reddening present in the field of this GGC. When we applied the PLZ
relations, assuming a metallicity of Z=0.0025 as in
\citetalias{alo15}, we could also see the effect of differential
extinction in the calculated distance moduli and color excesses (see
left panel in Fig.~\ref{fig_gc2dis}). This allowed us to perform an
analysis similar to that in \citetalias{alo15}, where we calculated
simultaneously extinction ratios and distance to this GGC from a
linear fit to the distance moduli and color excesses of its
RR~Lyrae. The zero-point of the fit is the distance modulus to this
GGC, corrected by extinction, while the slope of the fit is the
selective-to-total extinction ratio. Unfortunately, as was the case in
\citetalias{alo15}, the substantial reddening of 2MASS-GC~02 prevented
our detection of the RR~Lyrae in the bluer filters ($Z$ and $Y$), so
we could only perform our analysis using $J$, $H$, and $K_s$. Hence,
following \citetalias{alo15}, we performed an ordinary least-squares
bisector fit on the $\mu_{K_s}$ versus E($J-K_s$) and on $\mu_{K_s}$
versus E($H-K_s$) values obtained from the PLZ relations for the RRab
stars and we obtained the selective-to-total extinction ratios
$R_{K_s,H-K_s}=1.13\pm0.11$ and $R_{K_s,J-K_s}=0.55\pm0.08$, which
agree within the errors with those from \citetalias{alo15}. However,
we should note that while $R_{K_s,H-K_s}$ is slightly lower than in
our previous work, $R_{K_s,J-K_s}$ is instead slightly higher. The
mean of the distance moduli corrected by extinction obtained from both
fits is $\mu_0=13.9\pm0.3$ mag, which translates into the distance
given in Table~\ref{tab_gcdis}. While this distance differs from that
provided by the Harris catalog, it closely agrees with the distance
provided in \citetalias{alo15}, once the effects of the recalibration
of the PLZ relations are taken into consideration. The color excesses
reported in Table~\ref{tab_gcdis} for 2MASS-GC~02 are the mean of
those obtained to the individual RR~Lyrae, and they are similar to
those reported in \citetalias{alo15}. However, we note that these
average color excesses for 2MASS-GC~02 should be used with caution
because extinction toward this GGC is highly variable and changes
significantly over small regions.

For Terzan~10, we identified 7 RRab stars as cluster members following
the analysis in Sect.~\ref{sec_gc2pm}, as we can observe in
Fig.~\ref{fig_gc2} and \ref{fig_terzan10var}. They are the same 7 RRab
that we identified in \citetalias{alo15} as members. In the right
panel of Fig.~\ref{fig_gc2dis}, we show the distance moduli and color
excesses that we obtained after applying the PLZ relations, taking now
$Z_{Terzan10}$=0.0007; this is a lower metallicity than inferred in
\citetalias{alo15}, which we adopted for this work given the lower
iron content for this GGC reported in Table~\ref{tab_gcinput}. There
is indication of differential reddening, but it is less significant
than in 2MASS-GC~02. Importantly, the bulk of the RR~Lyrae in this GGC
does not seem to be suffering from it, with only a couple of these
sources (C24, C5) showing significant departures from the mean
reddening. So instead of performing a linear fit that would be heavily
affected by the extreme values of the color excesses of just these two
stars, we preferred to apply the ratios $R_{K_s,\lambda-K_s}$ from
\citet{alo17} to correct for extinction and calculate their
distances. The mean of these measurements is presented in
Table~\ref{tab_gcdis} as the distance to Terzan~10, and although it is
clearly off from the value given in the Harris catalog, this value
agrees with that provided in \citetalias{alo15} and also agrees with
recent measurements by \citet{ort19}.

\section{Conclusions}
\label{sec_con}
Using the VVV survey, we studied for the first time in the
near-infrared the variable stellar population over the entire field of
M~22, M~28, NGC~6569, and NGC~6441, four massive GGCs located toward
the Galactic bulge, whose corresponding metallicities span a wide
range of values. We also revisited the topic in 2MASS-GC~02 and
Terzan~10, which are two poorly populated and highly reddened GGCs we
already studied in the first paper of this series. The updated VIRAC2
database provided us with PMs and light curves for the stars located in
these GGCs. We defined a parameter that allows us to efficiently
discriminate the light curves provided by the VIRAC2 database and to
single out those which show clear signs of variability. We were able
to identify almost all of the RRab pulsators reported in other
catalogs of variable stars for these GGCs, except for the innermost
regions of the farthest clusters. Moreover, we were able to catalog
some other known RRc and Cepheid pulsators, and some other known
binary stars that show clear signs of variability in our VVV dataset,
and we identified some tens of new variable candidates. We were also
able to recover the vast majority of the variable candidates found for
2MASS-GC~02 and Terzan~10 in the first paper of this series, plus a
significant number of new candidates. We used the PMs that VIRAC2
provides to identify cluster members through a kNN classifier. We were
able to provide in this way the PMs for these GGCs, which agreed with
those provided by {\em Gaia} DR2, except for the most reddened
cluster, 2MASS-GC~02, where the VVV near-infrared observations provide
a more accurate result. Using their PMs, along with their positions in
the sky and in the CMD, we were able to select the variable stars that
belong to these GGCs as well. Since all of these clusters have a
significant number of RR~Lyrae, we used their tight, near-infrared PLZ
relations to calculate the distances and extinctions toward these
GGCs. We recalibrated previously used PLZ relations, obtaining in this
way a good agreement with those distances provided in the literature
and by {\em Gaia} DR2, except in the case of the Oosterhoff~III GGC
NGC~6441. Building on the methods described in this work, we plan to
extend the study of their variable stellar population to the other
GGCs located in the footprints of the VVV and the VVVX surveys.

\begin{acknowledgements}
J.A.-G., K.P.R., and S.R.A. acknowledge support from Fondecyt Regular
1201490. M.C. and D.M acknowledge support from Fondecyt Regular
1171273 and 1170121, respectively. J.A.-G., M.C., C.N., J.B., R.C.R.,
and F.G. are also supported by ANID – Millennium Science Initiative
Program – ICN12\_009 awarded to the Millennium Institute of
Astrophysics MAS. M.C., D.M. and D.G. are also supported by the BASAL
Center for Astrophysics and Associated Technologies (CATA) through
grant AFB-170002. C. E. F. L. acknowledges a post-doctoral fellowship
from the CNPq. F.G. also acknowledges support by CONICYT/ANID-PCHA
Doctorado Nacional 2017-21171485S. E.R.G. acknowledges support from
the Universidad Andr\'es Bello (UNAB) PhD scholarship
program. D.G. also acknowledges financial support from the Direcci\'on
de Investigaci\'on y Desarrollo de la Universidad de La Serena through
the Programa de Incentivo a la Investigaci\'on de Acad\'emicos
(PIA-DIDULS). R.A. also acknowledges support from Fondecyt
Iniciaci\'on 11171025.
\end{acknowledgements}

%
\bibliographystyle{aa} 
\bibliography{mybibtex} 
%




\end{document}